\documentclass[oldversion]{aa}  
\usepackage{graphicx,float}
\usepackage{latexsym,amsmath,amssymb}
\usepackage{natbib}
\usepackage{changebar}
\usepackage{subfigure}
\usepackage{txfonts}
\usepackage{rotating}
\usepackage{longtable, lscape}
\usepackage{hyperref}
\usepackage{enumerate}
\usepackage{color}

\usepackage{natbib}
\bibpunct{(}{)}{;}{a}{}{,}

\def\apjl{ApJ}%
\def\apjs{ApJS}%
\def\ao{Appl.~Opt.}%
\def\aap{A\&A}%
\begin{document}
   \title{A hard X-ray view of the soft excess in AGN}

   \subtitle{}

   \author{Rozenn Boissay
          \inst{1},
          Claudio Ricci
	\inst{2,3},
	and St\' ephane Paltani
          \inst{1}
          }

   \institute{ Department of Astronomy, University of Geneva, ch. d'\'Ecogia 16, 1290 Versoix, Switzerland
    \and Pontificia Universidad Cat\'olica de Chile, Instituto de Astrof\'isica, Casilla 306, Santiago 22, Chile \and
EMBIGGEN Anillo, Concepci\'on, Chile \\
             }
   \authorrunning{Rozenn Boissay}
   \titlerunning{A hard X-ray view of the soft excess in AGN}
    \date{Received; accepted}

 
 \abstract{An excess of X-ray emission below 1\,keV, called soft excess, is detected in a large fraction of Seyfert 1-1.5s. The origin of this feature remains debated, as several models have been suggested to explain it, including warm Comptonization and blurred ionized reflection. In order to constrain the origin of this component, we exploit the different behaviors of these models above 10\,keV. Ionized reflection covers a broad energy range, from the soft X-rays to the hard X-rays, while Comptonization drops very quickly in the soft X-rays. We present here the results of a study done on 102 Seyfert 1s (Sy 1.0, 1.2, 1.5 and NLSy1) from the \textit{Swift} \textit{BAT} 70-Month Hard X-ray Survey catalog. The joint spectral analysis of \textit{Swift}/\textit{BAT} and \textit{XMM-Newton} data allows a hard X-ray view of the soft excess that is present in about 80\% of the objects of our sample. We discuss how the soft-excess strength is linked to the reflection at high energy, to the photon index of the primary continuum and to the Eddington ratio. In particular, we find a positive dependence of the soft excess intensity on the Eddington ratio. We compare our results to simulations of blurred ionized-reflection models and show that they are in contradiction. By stacking both \textit{XMM-Newton} and \textit{Swift/\textit{BAT}} spectra per soft-excess strength, we see that the shape of reflection at hard X-rays stays constant when the soft excess varies, showing an absence of link between reflection and soft excess. We conclude that the ionized-reflection model as the origin of the soft excess is disadvantaged in favor of the warm Comptonization model in our sample of Seyfert 1s.

 }
   \keywords{accretion, accretion disks -- galaxies: active -- galaxies: nuclei -- galaxies: Seyferts -- X-rays: galaxies 
               }

   \maketitle
   
\section{Introduction}

The typical X-ray spectrum of a Seyfert 1 galaxy is composed of a cut-off power-law continuum, reflection features, low energy absorption, and often an excess in the soft X-ray below 1\,keV. 
The primary continuum power law is believed to stem from Comptonization of optical/UV photons from the accretion disk by energetic electrons in a hot (kT$\sim$100\,keV), optically-thin ($\tau\sim0.5$) corona \citep{Blandford1990, Zdziarski1995, Zdziarski1996, Krolik1999}. 
The reflection component is composed of a Compton hump peaking at $\sim$ 30\,keV and fluorescence lines, the most important of which is the Fe K$\alpha$ line around 6-7\,keV. This reflection can be due to reprocessing of the primary X-ray continuum on distant neutral material such as the molecular torus present in the unification model \citep{Antonucci1993, Jaffe2004, Meisenheimer2007, Raban2009} on the broad- and narrow-line regions \citep{Bianchi2008,Ponti2013} or in the accretion disk \citep{George1991, Matt1991}. 

While Seyfert 2 galaxies are generally highly absorbed, for example by the dusty torus requested by the unification model \citep{Antonucci1993}, Seyfert 1s are usually not absorbed or only lightly ($N_{\rm\,H}\leq10^{22} \text{ atoms cm}^{-2}$). When the absorbing material is photoionized, it is referred to as “warm absorber” \citep{George1998}. Absence of absorption allows us to see a soft X-ray emission in excess of the extrapolation of the hard X-ray continuum in many Seyfert 1s below 1\,keV. Discovered by \cite{Singh1985} and \cite{Arnaud1985} using {\it HEAO-I} and {\it EXOSAT} observations, this feature has since then been called {\it soft excess} (SE) and is detected in more than 50\% of Seyfert 1s \citep{Halpern1984,Turner:1989fk}. This fraction can reach 75\% to 90\% according to \cite{Piconcelli2005}, \cite{Bianchi2009}, and \cite{Scott2012}. 
The soft excess was first believed to originate in the inner part of the accretion flow \citep{Arnaud1985,Pounds1986}, and it has been modeled for a long time using a blackbody with temperatures of $\sim0.1-0.2$\,keV. This model has been ruled out since a standard accretion disk around a supermassive black hole can not lead to such high temperatures \citep{Shakura1973}. Moreover, the temperature of the blackbody used to model the soft excess has been found to be independent of the black-hole mass, unlike what is expected from a standard accretion disk \citep{Gierlinski2004}.

A possible explanation for the soft excess is the ``warm" Comptonization scenario: UV seed photons from the disk are upscattered by a Comptonizing corona cooler and optically thicker than the hot corona responsible for the primary X-ray emission. This model has been successfully applied to NGC 5548 \citep{Magdziarz1998}, RE J1034+396 \citep{Middleton2009},  RX J0136.9$-$3510 \citep{Jin2009},  Ark 120 \citep{Matt2014}, and 1H 0419$-$577 \citep{DiGesu2014}. This Comptonization model is supported by the existence of similarities between the spectral shape \citep{Walter1993} and the variability of the optical/UV and soft X-ray emission \citep{Edelson1996}. In a more recent work, \citet{Mehdipour2011} found, in the framework of a multi-wavelength campaign on Mrk\,509, a strong correlation between fluxes in the optical/UV and soft X-ray bands, as it would be expected if the soft excess was due to warm Comptonization of the seed photons from the disk. \cite{Pop2013} used ten simultaneous \textit{\textit{XMM-Newton}} and \textit{INTEGRAL} observations of  Mrk 509 to find that a hot ($kT\sim100$\,keV), optically-thin ($\tau\sim0.5$) corona is producing the primary continuum and that the soft excess can be modeled well by a warm (kT $\sim$ 1\,keV), optically-thick ($\tau\sim$ 10-20) plasma. A warm Comptonization model is also confirmed in Mrk 509 by considering the excess variability in the soft-excess flux on long time scales and variability properties in general \citep{Boissay2014}.
Using five observations of Mrk 509 with \textit{Suzaku}, \cite{Noda2011} showed that the fast variability seen in hard X-rays is not present in the soft X-ray excess. This differential variability rules out a small reflector origin for the soft excess in Mrk 509, similar to what has been shown by \cite{Boissay2014}, because the soft X-rays should closely follow the hard X-ray variability in the relativistic ionized reflection models.

Since it is difficult for the Comptonization hypothesis to explain the consistency of the soft-excess feature over a wide range of black-hole masses $M_{\rm\,BH}$, another possible explanation for the soft excess, which is blurred ionized reflection, has been invoked. In this scenario, the emission lines produced in the inner part of the ionized disk are blurred by the proximity of the supermassive black hole (SMBH). 
\cite{Crummy2006} successfully applied such an ionized-reflection model (\texttt{reflionx}; \citealt{Ross2005}) on a sample of PG quasars to explain the soft excess. \cite{Zoghbi2008} used a similar approach for Mrk 478 and EXO 1346.2+2645. This model has been used to explain the spectral shape, as well as the variability, in MCG$-$6$-$30$-$15 \citep{Vaughan2004} and NGC 4051 \citep{Ponti2006}. 
\cite{Walton2013} used a sample of 25 bare AGN observed with \textit{Suzaku} to test the robustness of the reflection interpretation by reproducing the broad-band spectra using the \texttt{reflionx} model (as done by \citealt{Crummy2006}) and to constrain the black hole spin in many objects of the sample.
More advanced blurred ionized-reflection models have been created in recent years, such as \texttt{relxill}, developed by \cite{Garcia2014} and \cite{Dauser2014}, which merges the angle-dependent reflection model \texttt{xillver} with a relativistic blurring model \texttt{relline}. This \texttt{relxill} model has already been used to explain the soft excess in Mrk 335 \citep{Parker2014} and in SWIFT\,J2127.4+5654 \citep{Marinucci2014}. \cite{Vasudevan2014} showed, using \textit{Swift/BAT} observations and \textit{XMM-Newton/NuSTAR} simulations, that a correlation between the reflection and the soft-excess strength is expected if this feature is due to blurred ionized reflection. Parameters obtained when  fitting spectra with ionized-reflection models are often extreme. A maximally rotating supermassive black hole and a very steep emissivity are required in most cases \citep{Fabian2004,Crummy2006}, as well as fine-tuning of the ionization of the disk \citep{Done2007}. These parameters, which are similar to those obtained when modeling the broad component of the Fe K$\alpha$ line \citep{Fabian:2005mz}, show that in this scenario most of the reprocessed radiation is produced very close to the event horizon.
Time delays between iron lines and direct X-ray continuum have been detected in several objects (e.g., 1H 0707$-$495, \citealt{Fabian2009}; NGC 4151, \citealt{Zoghbi2012}; MCG$-$5$-$23$-$16 and NGC 7314, \citealt{Zoghbi2013}; Ark 564 and Mrk 335, \citealt{Kara2013}). Soft X-ray reverberation lags have been detected in a considerable sample of objects \citep{Cackett2013,DeMarco2011,DeMarco2013} and are interpreted as a signature of reverberation from the disk, which would support the reflection origin for the soft excess. However, if soft lags are detected on short time scales, the soft X-ray emission can lead the hard band on long time scales, as in PG\,1244+026 \citep{Gardner2014}. In this case, the soft excess is interpreted as a combination of intrinsic fluctuations propagating down through the accretion flow giving the soft lead and reflection of the hard X-ray emission giving the soft lag. 

Using a sample of six bare Seyfert galaxies, \cite{Patrick2011} modeled the \textit{Suzaku} broad-band spectra with both ionized reflection and additional Compton scattering components. Combining the two models to reproduce the soft excess provided better fitting results with lower values of the spin and emissivity index. Other works presenting direct tests of the ionized reflection versus warm Comptonization include, for example, analyses from \cite{Lohfink2013} and \cite{Noda2013}. \cite{Lohfink2013} used archival \textit{XMM-Newton} and \textit{Suzaku} observations of Mrk 841 to test different models for the origin of the soft excess in a multi-epoch fitting procedure. \cite{Noda2013} fit \textit{Suzaku} spectra of five AGN with different models to reproduce the soft excess and concluded that the thermal Comptonization model reproduces the data better.

The soft excess could also be the signature of strong, relativistically smeared, partially ionized absorption in a wind from the inner disk, as proposed for PG 1211+143 by  \cite{Gierlinski2004}. In the case of a totally covering absorber, extreme parameters are needed to reproduce the observed soft excess, as in the case of blurred ionized-reflection models. For example, a very high smearing velocity ($v \sim 0.3 c$) is required, but such a velocity is difficult to reach considering radiatively driven accretion disk winds \citep{Schurch2007, Schurch2009}. 
We do not further discuss this model in this paper.

This papers aims to test blurred ionized-reflection models as the origin of the soft excess, in particular in the lamp-post configuration, via a spectral model-independent analysis. Different scenarii of the origin of the soft excess imply  different behaviors in the soft and hard X-ray bands, since ionized reflection covers a broad energy range, while Comptonization drops very quickly in the soft X-rays. Because we want to get good constraints on the hard X-ray reflection measurements, we consider sources detected by the \textit{Swift/\textit{BAT}}  instrument and combine these observations with data from the \textit{XMM-Newton} satellite to get information about the soft excess. We carry out broad-band (0.5-100\,keV) spectral analysis of the sources and explore the relations between the parameters resulting from the fitting procedure. We compare the reflection and soft-excess parameters with those obtained from simulations in a lamp-post configuration. We also study the differences between soft and hard X-ray emission by stacking the \textit{XMM-Newton} and \textit{BAT} spectra according to their soft-excess strengths. 

\section{Sample and data analysis}
\label{2}

We use a sample of sources from the \textit{Swift} \citep{Gehrels2004} \textit{BAT} 70-Month Hard X-ray Survey catalog \citep{Baumgartner2013}, which contains 1210 sources including 292 Seyfert 1s. By cross-correlating the \textit{Swift}/\textit{BAT} catalog with the Veron catalog \citep{Veron2010}, we selected only narrow-line Seyfert 1s (NLSy1s) and Seyfert 1s - 1.5s (Sy1s - 1.5s). Our sample is in this way composed of 102 sources, divided in 37 Sy1.0s, 18 Sy1.2s, 35 Sy1.5s, and 12 NLSy1s.  We selected sources that have been observed with the \textit{XMM-Newton} satellite \citep{Jansen2001} as pointed observations with a minimum exposure time of 5 ks (see Table \ref{tab:epic_info}). We selected sources with a flux higher than $10^{-11} \text{ergs}/\text{s}/\text{cm}^{2}$ in the 14-195\,keV band. \textit{BAT} spectra \citep{Barthelmy2005} have been obtained from the \textit{Swift}/\textit{BAT} website \footnote{\url{http://Swift.gsfc.nasa.gov/results/bs70mon/}}. 

To study the soft X-ray emission of these objects, we used \textit{XMM-Newton} observations. In the case of multiple observations for the same source, we considered the observation with the longest exposure time (see Table \ref{tab:epic_info}). Data was reduced on the original data files using the \textit{XMM-Newton} Standard Analysis Software (SAS v12.0.1 - \citealt{Gabriel2004}) considering both the \textit{PN} \citep{Struder2001} and \textit{MOS} \citep{Turner2001} spectra of each source. Events were filtered using \#XMMEA\_EP and \#XMMEA\_EM, for \textit{PN} and \textit{MOS} cameras, respectively. Single and double events were selected for extracting \textit{PN} spectra and single, double, triple, and quadruple events were considered for \textit{MOS}. The data were screened for any increased flux of background particles. Spectra were extracted from a circular region of 30 arcsec centered on the source. The background was extracted from a nearby source-free region of 40 arcsec in the same CCD as the source.
The \textit{XMM-Newton} data showed evidence of significant pile-up in some objects (see Table\,\ref{tab:epic_info}), for which the extraction region used was an annulus of inner radius 15 arcsec instead of a simple circle. Response matrices were generated for each source spectrum using the SAS \textit{arfgen} and \textit{rmfgen} tasks.

\section{Spectral analysis}
\label{3}
In this section, we explain the broad-band spectral analysis procedure of all the sources of our sample between 0.5 and 100\,keV.
More precisely, we fit the 3-100\,keV data with a phenomenological model and then quantify the soft excess below 2 keV with respect to this model. We present different fitting models, taking reprocessing into account that is mainly due to distant and/or local reflectors.

\subsection{Spectral fitting procedure}
\label{3.1}
\subsubsection{Reprocessing mainly due to a distant reflector}
\label{3.1.1}

The spectral analysis of our sample is performed using the general X-ray spectral-fitting program XSPEC (version 12.8.1, \citealt{Arnaud1996}).
To determine specific parameters for each object of our sample, we analyzed  \textit{PN} and \textit{MOS} spectra from the \textit{XMM-Newton} satellite and \textit{BAT} spectra from \textit{Swift}. 

We first chose to fit \textit{PN}/\textit{MOS} and \textit{BAT} spectra jointly in an energy band in which there is little contamination from the soft excess or absorption (from 3 to 100\,keV) with a cut-off power law and a neutral reflection component using the \texttt{pexmon} model \citep{Nandra2007} from XSPEC. The \texttt{pexmon} model is similar to the \texttt{pexrav} model \citep{Magdziarz1995}, but includes  Fe K$\alpha$, Fe K$\beta$, and Ni K$\alpha$ lines and the Fe K$\alpha$ Compton shoulder \citep{George1991}. We use the \texttt{pexmon} model to take the reflection component into account, the cut-off power-law component being modeled by \texttt{cutoffpl}. We also take Galactic absorption (\texttt{abs$_{\texttt{Gal}}$}) into account, fixing the value of $N_{\rm\,H}^{G}$ to the one found in the literature \citep{Dickey:1990fj} (see Table \ref{tab:epic_info}). 

The \texttt{pexmon} model includes the narrow iron lines produced by reflection from distant neutral material, such as the broad-line region (e.g., \citealt{Bianchi2008}), the molecular torus (e.g., \citealt{Shu2011,Ricci2014}), or the region between the torus and the broad-line region \citep{Gandhi2015}. However, assuming that the primary X-ray continuum is also reprocessed in the innermost part of the accretion disk, these iron lines should be broadened by Doppler motion and relativistic effects (e.g., \citealt{Fabian1989}). To consider these effects, we allow broadening of the iron line from \texttt{pexmon}, by convolving the \texttt{pexmon} model with a Gaussian smoothing model \texttt{gsmooth} ($\sigma$ can vary between 0.0 and 0.5\,keV).

\textit{XMM-Newton} and \textit{Swift}/\textit{BAT} observations are not simultaneous because \textit{BAT} spectra are integrated over 70 months, so variability in flux can occur between \textit{XMM-Newton} and \textit{Swift}/\textit{BAT} spectra. To correct the flux variability during the fitting procedure, we multiply the cut-off power-law of our model by a cross-calibration factor $f$. The \texttt{pexmon} component parameters are fixed to the same value for \textit{XMM-Newton} and \textit{BAT}. When doing so, we consider that the reflection component, assumed to be mainly due to a distant reflector, does not vary on the \textit{BAT} time scale and that only the primary continuum does. The cross-calibration factor $f$ is fixed to 1 for \textit{BAT} spectra and left free between 0.7 and 1.3 for \textit{XMM-Newton} spectra ($f$ is found to be consistent with being equal for \textit{PN} and \textit{MOS} spectra). Using this $f$ factor, we constrain the fit sufficiently while taking the flux variability of the primary continuum between \textit{XMM-Newton} and \textit{BAT} into account.

The resulting model for the fit between 3 and 100\,keV is

\begin{center}
\texttt{abs$_{\texttt{Gal}}$\,(f$\times$cutoffpl+gsmooth*pexmon)}.
\end{center}

When fitting the \textit{XMM-Newton} and \textit{Swift}/\textit{BAT} spectra between 3 and 100\,keV, we keep the inclination angle of the disk to its default value of 60$^{\circ}$. Abundances in \texttt{pexmon} are free to vary between 0.3 and 3 (relative to solar), in order to properly model the iron line as well as the Compton hump. The high-energy cut-off can vary from 100 to 500\,keV. The reflection factor $R$, which is the strength of the reflection component relative to what is expected from a slab subtending $2\pi$ solid angle, can have values between 0 and 100.
We find the best fit parameter values using the $\chi^{2}$ statistics of XSPEC. 

To study the soft X-ray emission of the objects, we fix the parameters of the model described previously to the values found during the fit between 3 and 100\,keV and add \textit{XMM-Newton} data in the 0.5 to 3\,keV energy band. We remove the \texttt{gsmooth} component from our model to avoid spurious features at low energy (arising from the convolution at the edge of the spectrum). Suppressing this component will just slightly degrade the $\chi^{2}$ value and give an advantage in computation time.

Since we want this analysis to be model independent, we do not favor any hypothesis here for the origin of the soft excess. To measure the strength of the soft excess, we use a Bremsstrahlung model that provides a good phenomenological representation of its smooth spectral shape. We then find the best-fit parameter values using the $\chi^{2}$ statistics, fitting the spectra between 0.5 and 100\,keV.  In the majority of the objects of our sample, the soft excess is fit better by two Bremsstrahlung models than by a single one with plasma temperatures between 0 and 1\,keV.

We also evaluate the presence of absorption from a cold or ionized medium (\texttt{abs$_{\texttt{cold/warm}}$}). The presence of a single or double warm absorber (WA), modeled by XSTAR, and/or of a cold absorber, modeled by \texttt{zwabs}, is verified by an F test between the model that takes only the soft excess into
account and the one that includes the absorber.

The resulting broad-band-fitting model is

\begin{center}
\texttt{abs$_{\texttt{Gal}}$$\times$abs$_{\texttt{cold/warm}}$\,(bremss+bremss+f$\times$cutoffpl+pexmon)}.
\end{center}

\subsubsection{Reprocessing mainly due to a local reflector}
\label{3.1.2}

In Sect. \ref{3.1.1}, we considered that the reflection component is mainly due to a distant reflector, which is expressed by the cross-calibration being applied to the power law and not to the \texttt{pexmon} component in the fitting procedure. We also want to study the case where the reflection is mainly due to a local reflector.
The model used for this hypothesis to fit the 3 to 100\,keV band is

\begin{center}
\texttt{abs$_{\texttt{Gal}}$$\times$f\,(cutoffpl+gsmooth*pexmon)}.
\end{center}

Applying this $f$ cross-calibration factor on the \texttt{pexmon} component this time allows us to consider that the reflection responds with very little lag to the continuum. We fit this new model on our objects with the same ranges of parameters as described in Sect. \ref{3.1.1}. The fitting procedure between 0.5 and 100\,keV is the same as the one described in Sect. \ref{3.1.1} with the model

\begin{center}
\texttt{abs$_{\texttt{Gal}}$$\times$abs$_{\texttt{cold/warm}}$\,(bremss+bremss+f\,(cutoffpl+pexmon))}.
\end{center}


\subsubsection{Comparison between local and distant reflection models}
\label{3.1.3}
\begin{figure} [!b]
\resizebox{\hsize}{!}{\includegraphics[trim = 10mm 60mm 10mm 70mm, clip, angle=0]{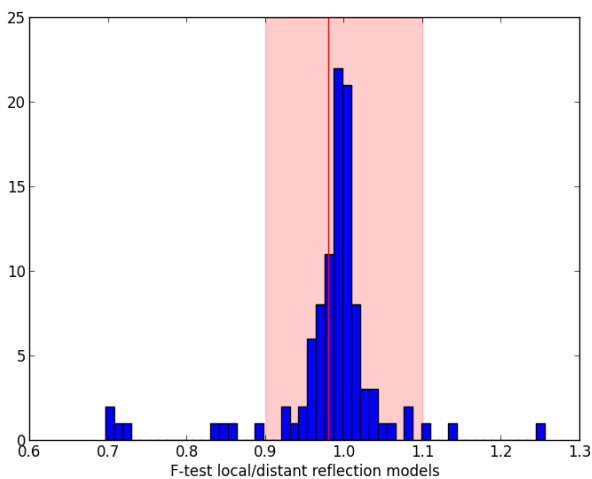}} 
\caption{Distribution of the F test results between the local reflection model and the distant reflection model used to fit each object. The red area is the acceptance zone of the null hypothesis. The average F value, represented by the red line, is inside the acceptance zone.}
\label{ChiComp}
\end{figure}

For each object of our sample, we perform a statistical F test between models of local and distant reflection. We can see in Fig. \ref{ChiComp} that, considering the theoretical F values range from the Fisher table for which we can accept the null hypothesis at a 95\% significance level (see red area in Fig. \ref{ChiComp}), only 8\% of our objects are better fit by the local reflection model and 3\% by the distant reflection model. The average value of F being less than 1 ($\text{F}_{\text{mean}}$=0.98) shows a little preference for the local-reflection model to fit our data, but since $\text{F}_{\text{mean}}$ is inside the acceptance zone, this preference is not significant. We therefore cannot distinguish between the local-reflection and the distant-reflection models here.

The X-ray spectra of our Seyfert galaxies should present iron emission lines with two components: the narrow line coming from reflection on the dusty torus or the broad-line region and on the broad line thought to be produced in the inner accretion disk. The total reflection factor includes the reflection strengths of both components. In the hypothesis where the soft excess is due to ionized reflection, the broad line is expected to be more important than the narrow one, which is why we use a \texttt{gsmooth} model to allow the broadening of the iron line from the \texttt{pexmon} model. Furthermore, we have just seen that changing the input of the cross-calibration factor to account for local and distant reflections does not allow us to distinguish the models. However, to check that we do not miss any information from the narrow iron lines, we tested a fitting procedure with two \texttt{pexmon} models on 10\% of the objects of our sample. One simple \texttt{pexmon} model was used to reproduce the narrow iron line, and the second one, convolved with a \texttt{gsmooth} model, represents the broad component. The sum of the reflection fractions from the two \texttt{pexmon} models is equal to the reflection fraction measured with our one-\texttt{pexmon} fitting procedure. This test was done on ten objects
for which we have good signal-to-noise data. As the results are consistent between the two approaches and as the reflection 
factor is better constrained in our initial fitting procedure, we chose to keep this one-reflection
fitting process to model our data.

\subsection{Soft excess and absorption properties} 

The results obtained by our spectral analysis, which is based on the hypothesis of a reflection due mainly to distant material, are presented in Table \ref{tab:fit}. We find that 79 objects (i.e., 80\% of our sample) present a soft excess (SE). Among these 79 objects showing a soft excess, 42 do not show the presence of absorption and 37 are hardly absorbed by cold or warm material. 

The 24 other objects show the presence of a stronger absorption by a cold or warm absorber with a column density value $N_{\rm\,H}\geq10^{22} \text{ atoms cm}^{-2}$.
These absorbers prevent a good measurement of an eventual soft excess. Twelve of these strongly absorbed objects present a complex absorption and are labeled in Table \ref{tab:fit}: 
\begin{enumerate}[(a)]
\item ESO 323$-$77: \cite{Miniutti2014} used multi-epoch spectra from 2006 to 2013 to show that the absorption is due to a clumpy torus, broad-line regions, and two warm absorbers.
\item EXO 055620$-$3820.2: \cite{Turner1996} found by using \textit{ASCA} observations that the continuum of the source is attenuated by an ionized absorber either fully or partially covering the X-ray source. 
\item IC 4329A: \cite{Steenbrugge2005} used \textit{XMM-Newton} observations to show that the absorber is composed of seven different absorbing systems.
\item MCG$-$6$-$30$-$15: \cite{Miyakawa2012} fit \textit{Suzaku} data considering absorbers with only a variable covering factor. 
\item Mrk 1040: \cite{Reynolds1995} found that the soft spectral complexity visible in \textit{ASCA} observations could be either explained by a soft excess and by intrinsic absorption or by a complex absorber.
\item Mrk 6: \cite{Mingo2011} showed that the variable absorption visible in \textit{XMM-Newton} and \textit{Chandra} observations are probably caused by a clump of gas close to the central AGN in our line of sight.
\item NGC 3227: \cite{Beuchert2014} used \textit{Suzaku} and \textit{Swift} observations of a 2008 eclipse event to characterize the variable-density absorption, probably caused by a filamentary, partially covering, and moderately ionized cloud.
\item NGC 3516: \cite{Huerta2014} showed that this object is absorbed by four warm absorbers according to nine \textit{XMM-Newton} and \textit{Chandra} observations.
\item NGC 3783: \cite{Brenneman2011} used \textit{Suzaku} observations to show the presence of a multicomponent warm absorber.
\item NGC 4051: \cite{Pounds2013} used a \textit{XMM-Newton} observation to find a fast, highly ionized wind, launched from the vicinity of the supermassive
black hole, that shocked against the interstellar medium (ISM). They speculate that the warm absorbers often observed in AGN spectra result from an accumulation of such shocked winds. 
\item NGC 4151: \cite{Wang2011} found, in a Chandra observation, emission features in soft X-rays that are consistent with blended brighter O VII, O VIII, and Ne IX lines. They also found low and high ionization spectral components that are consistent with warm absorbers.
\item UGC 3142: \cite{Ricci2010} used \textit{XMM-Newton}, \textit{Swift}, and INTEGRAL data for the spectral analysis. This object is absorbed by two layers of neutral material. 
\end{enumerate}

We decide not to include these 23 absorbed objects in our analysis in order to have a clean measurement of the soft-excess intensity.

\subsection{Soft-excess strength}
For the following analysis, we  keep only the 79 objects that show the presence of soft excess and  that are absorbed by a material with a $N_{\rm\,H}$ smaller than $10^{22} \text{ atoms cm}^{-2}$.
For each of these objects, we define the strength of the soft excess \textit{q} as the ratio between the flux of the soft excess (i.e., the flux of the Bremsstrahlung models between 0.5 and 2\,keV) and the extrapolated flux of the continuum between 0.5 and 2\,keV. This definition differs from the one used in \cite{Vasudevan2014}, as discussed in Sect. \ref{4.2}.

\section{Relation between reflection and soft-excess strength}
\label{4}
In this section, we compare the reflection and soft-excess strength from data with those obtained from simulations of blurred ionized reflection. These simulations, similar to the ones of \cite{Vasudevan2014}, are performed using the lamp-post configuration of the \texttt{relxill} model of \cite{Garcia2014} and \cite{Dauser2014}, changing the height of the source that controls the emissivity index and the reflection fraction and changing the mass accretion rate that controls the inner disk ionization. The resulting simulated spectra are then fit with the same phenomenological model than real data to compare the distributions of soft-excess strength versus reflection strength from the real and simulated data. 

\subsection{$R$ vs $q$ in the sample}
\label{4.1}
\begin{figure} [!b]
\begin{minipage}[c]{1.\linewidth}
\resizebox{\hsize}{!}{\includegraphics[trim = 10mm 60mm 10mm 70mm, clip, angle=0]{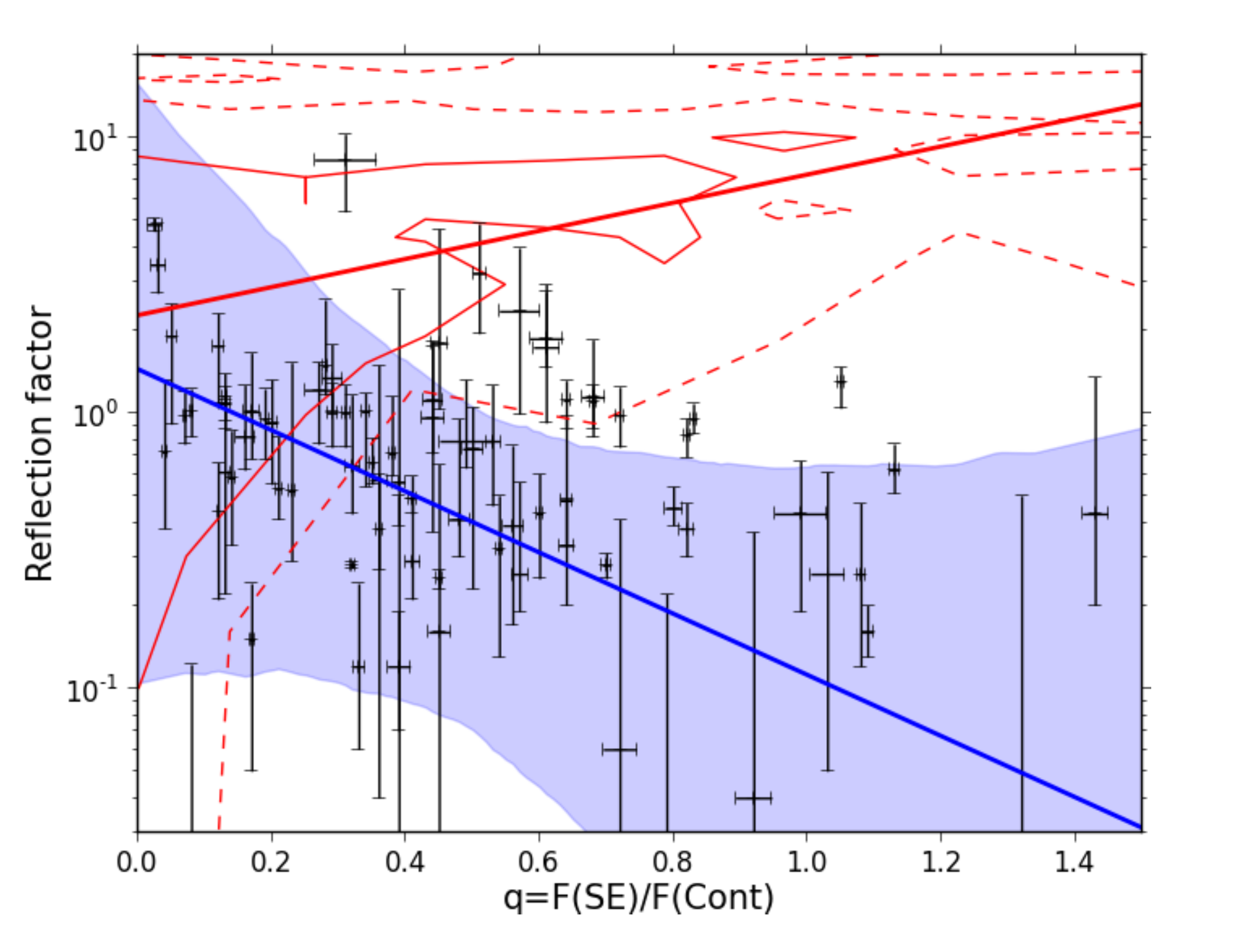}} 
\end{minipage}
\begin{minipage}[c]{1.\linewidth}
\resizebox{\hsize}{!}{\includegraphics[trim = 10mm 60mm 10mm 70mm, clip, angle=0]{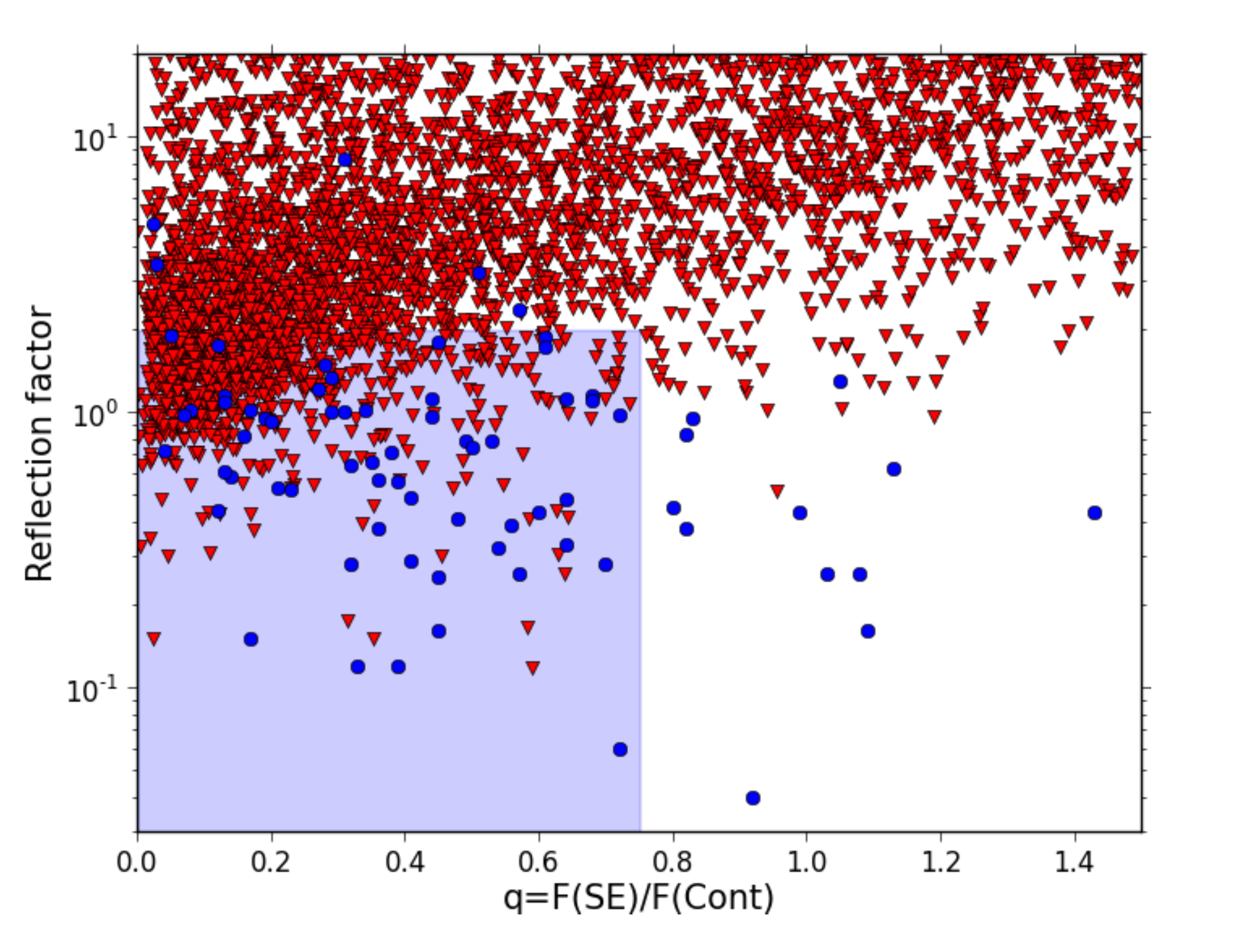}} 
\end{minipage}
\caption{Reflection factor $R$ as a function of the strength of the soft excess $q$ (in the case of distant reflection). Top panel: Linear regression performed using the \texttt{linmix\_err} procedure is represented by the blue line, and 99.9\% CI is given by the blue contour. The red line represents the positive correlation expected if the soft excess is due to ionized reflection (result of \texttt{relxilllp\_ion} simulations, see Sect. \ref{4.2}). Density contours of the simulations are plotted in red (solid line: 68\%; dashed line: 90\%). Bottom panel: Blue points represent our 79 objects showing a soft excess. They are mostly contained (75\%) in the blue box ($R<2.0$ and $q<0.75$). Red triangles are \texttt{relxilllp\_ion} simulation results.}
\label{RvsSE}
\end{figure}

To study the evolution of the soft excess with reflection, we plot, in top panel of Fig. \ref{RvsSE}, the reflection factor measured when fitting \textit{XMM-Newton} (\textit{EPIC PN} and \textit{MOS}) and \textit{Swift/BAT} spectra as a function of the soft-excess strength $q$. In the case of distant reflection (see Sect. \ref{3.1.1}), we find an anti-correlation characterized by a Spearman coefficient of r=-0.33 and a null-hypothesis probability of about 0.3\%. We perform a bootstrap on the data, finding a 99.9\% confidence interval (CI) for the Spearman correlation coefficient of $-0.62 \leq r \leq 0.01$ and a probability of having a negative correlation of 99.8\%. 

Our sample includes NLSy1s, objects that often present a strong soft excess and a steep spectrum (e.g., \citealt{Vaughan1999,Haba2008,Done2013}). If we do not include the ten NLSy1s in our analysis, we still find a correlation with a Spearman coefficient of $r=-0.31$, a null-hypothesis probability of about 0.7\% and a probability of 99\% of having a negative correlation between $R$ and $q$. 

To look at the effect of the warm absorber in the objects of our sample, we only consider the 42 objects showing a soft excess without absorber. Spearman statistics give a correlation coefficient of r=-0.37 with a null-hypothesis probability of 2\% and the probability of having a negative correlation of 98\%. The anti-correlation between $R$ and $q$ found with the entire sample of objects showing a soft excess with or without absorber still exists when considering only objects without absorber. In this case, the anti-correlation is less significant because it is based on about half of the sample.

To consider errors in both x and y axes, as well as intrinsic scatter, we performed a linear regression with a Bayesian approach using an IDL procedure called \texttt{linmix\_err} \citep{Kelly2007}. 
The linear regression process results in the relation: 
\begin{equation}
R=-0.72\,_{-0.21}^{+0.28} \times q+1.01\,_{-0.11}^{+0.18}
.\end{equation}
The intrinsic scatter of $R$ is 0.37.  The Markov Chains Monte Carlo (MCMC) created by the IDL procedure allowed us to plot the 99.9\% CI for this linear regression (blue contour in top panel of Fig. \ref{RvsSE}).

In the case of local reflection (see Sect. \ref{3.1.2}), Spearman statistics give a correlation coefficient of r=-0.48 between $R$ and $q$ and a null-hypothesis probability of 0.06\%. The linear regression performed with \texttt{linmix\_err} results in the relation:

\begin{equation}
R=-0.54\,_{-0.12}^{+0.09} \times q+0.94\,_{-0.01}^{+0.20}
.\end{equation}
The intrinsic scatter of $R$ is 0.24.

If we only consider NLSy1 galaxies, Spearman statistics give a correlation coefficient of r=-0.36, a null-hypothesis probability of 31\% and a probability of 87\% of a negative correlation. The significance of this result is still quite strong considering that our sample contains only ten NLSy1s for which we can measure the soft excess.


\subsection{$R$ vs $q$ in simulations of ionized reflection}
\label{4.2}

\cite{Vasudevan2014} performed $\sim$2400 \textit{XMM--Newton} and \textit{NuSTAR} simulations of blurred ionized reflection, using \texttt{reflionx} and \texttt{kdblur} and taking neutral reflection with \texttt{pexrav} into account. They used a wide range of values for the iron abundance, the photon index, the ionization, and the ratio between the normalizations of \texttt{pexrav} and \texttt{reflionx}. They fixed the emissivity index and the inclination to intermediate values and fixed the high energy cut-off to $10^{6}$\,keV. The strength of the hard excess $R$ was then measured by a neutral reflection model \texttt{pexrav} (the iron line being modeled by a Gaussian component) and the soft excess was modeled by a blackbody. They fixed the normalizations of each component of this model, in order to constrain the blackbody temperature and the Gaussian line energy and width. They defined the soft-excess strength as being the ratio between the luminosity from the blackbody between 0.4 and 3\,keV and the power-law luminosity between 1.5 and 6\,keV. Their simulations predict the existence of a correlation between $R$ and the soft-excess strength.

We want to determine the expected relation between the reflection factor and the soft-excess strength in the case of ionized reflection, similar to what was done by \cite{Vasudevan2014}. But since we want to use a more recent ionized-reflection model and our spectral fitting procedure differs from the one of \cite{Vasudevan2014}, we performed simulations using the \texttt{relxilllp\_ion} model.
The \texttt{relxilllp\_ion} model \citep{Dauser2014} is similar to the \texttt{relxill} model, but adapted for the lamp-post geometry. 
By simulating this lamp-post configuration, we assumed an explicit link between the smearing and the amount of reflection. For example, for objects, such as 1H 0707$-$495, that have a high emissivity index and are thus strongly smeared, the compact source is thought to be confined to a small region around the rotation axis and close to the black hole \citep{Fabian2012}. The very small height $h$ required predicts a very large reflection fraction $R$ (see the relation between $R$ and $h$ plotted in Figure 2 of \citealt{Dauser2014}).
The ionization of the accretion disk depends on the mass accretion rate $\dot{m}$ and on the density profile of the disk. The density structure assumed in a hydrostatic $\alpha$ disk \citep{Shakura1973} is not always appropriate. In particular, it has been shown that the accretion disk cannot be in hydrostatic equilibrium if the soft excess is made via reflection \citep{Done2007}. In the \texttt{relxilllp\_ion} model, the gas in the atmosphere of the accretion disk is assumed to be at a constant density \citep{Garcia2014}.

We simulated \textit{Swift}/\textit{BAT} and \textit{XMM-Newton}/\textit{PN} spectra with a signal-to-noise ratio comparable to our data (time exposure of 10 ks for \textit{XMM-Newton}/\textit{PN} and 1 Ms for \textit{Swift}/\textit{BAT}). We performed $\sim$3500 simulations with different values of \texttt{relxilllp\_ion} parameters for the reflection factor $R_{rel}$ (from 0 to 100, values that are reachable in a lamp-post configuration for a source very close to the black hole, see Fig. 5 in \citealt{Miniutti2004}), the photon index $\Gamma_{rel}$ (from 1.4 to 2.5), the ionization parameter at the inner edge of the disk ($log(\xi)$ between 0.0 and 4.7, as allowed by the model), the abundances $A_{Fe}$ (from 0.3 to 3.0 in solar unit), the height $h$ of the source (from 1 to 6 $r_{g}$), the inner radius $R_{in}$ (from 1 to 50 $r_{g}$), the density index (from 0 to 4), the mass accretion rate $\dot{m}$ (from 0.01 to 5), and the high-energy cut-off $Ec_{rel}$ (from 100 to 300\,keV). 
To account for the narrow component of the iron line in our simulated spectra, we added a \texttt{pexmon} component to our \texttt{relxilllp\_ion} model so as to reproduce the neutral reflection on the dusty torus and broad-line regions, with a reflection factor $R_{narrow}$ varying between 0 and 1. 
Considering only Seyfert 1s and according to the unification model, we performed simulations for different inclination values $\theta$ between 0$^{\circ}$ and 60$^{\circ}$, giving the weight $sin(\theta)$ to each simulation to take the probability of having such an inclination
into account.

We fit our simulated \texttt{relxilllp\_ion} spectra by using the fitting procedure explained in Sect. \ref{3.1}, in order to measure the photon index and the reflection from the \texttt{pexmon} model, as well as the value of the soft-excess strength $q$ and to be able to compare them to the parameters obtained when fitting the real data. We performed a linear regression 
(red solid line in top panel of Fig. \ref{RvsSE}) and plotted density contours of the simulated data (red contours in top panel of Fig. \ref{RvsSE}; solid line: 68\%, dashed line: 90\%). 
Similar to \cite{Vasudevan2014}, we expected to find a positive correlation between the reflection factor $R$ and the soft-excess strength $q$, if ionized reflection model is the explanation of the soft excess, even if the simulations and the soft-excess strength definition are slightly different. 
We found such a positive correlation with our simulations, assuming all \texttt{relxilllp\_ion} parameters are uncorrelated.
Spearman statistics give a correlation coefficient between $R$ and $q$ expected for ionized-reflection model of r=0.55 and a negligible null-hypothesis probability. The expected relation is

\begin{equation}
R=6.17\,_{-0.20}^{+0.19} \times q+2.80\,_{-0.11}^{+0.11}
.\end{equation}

The bottom panel of Fig. \ref{RvsSE} shows the reflection factor $R$ measured by \texttt{pexmon} as a function of the soft-excess strength $q$ for real and simulated data. Blue points represent the real data, i.e. results of the fitting of the objects of our sample, and red triangles are the parameters obtained by simulating \texttt{relxilllp\_ion} models (red contour in top panel of Fig. \ref{RvsSE} has been derived from these red triangles). As we can see in bottom panel of Fig. \ref{RvsSE}, 75\% of the blue data points are contained in the blue box ($R<2.0$ and $q<0.75$). \texttt{Relxilllp\_ion} simulations result in similar parameters, but also in higher reflection factors and higher soft-excess strengths. Twenty-five percent of our simulated objects are in the blue box, and 75\% of them are outside the blue box. We do not observe any high-$R$ objects, unlike what is expected with blurred ionized-reflection models (with the \texttt{relxilllp\_ion} model in our simulations, and with the \texttt{reflionx} model in simulations from \citealt{Vasudevan2014}).

\section{Evolution of the photon index of the primary continuum}
\label{5}
In this section, we explore the relations between the photon index of the primary continuum and the reflection and soft-excess strength. We compare the parameters derived from data analysis with those obtained from simulations of lamp-post configuration. 

\subsection{Soft excess and photon index}
\label{5.1}

\begin{figure} [!b]
\resizebox{\hsize}{!}{\includegraphics[trim = 10mm 60mm 10mm 70mm, clip, angle=0]{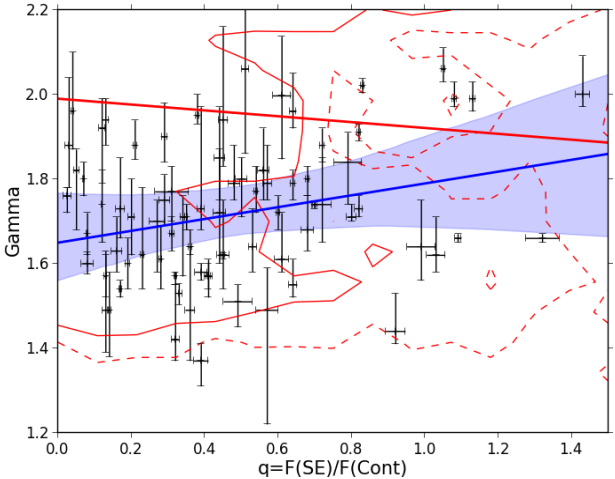}} 
\caption{Strength of the soft excess against the power-law slope obtained with \textit{XMM-Newton} and \textit{BAT} data (in the case of distant reflection). CI is represented by the blue contour. The red line represents the relation expected in the case of ionized reflection; the red contours are the result of \texttt{relxilllp\_ion} simulations (solid line: 68\%, dashed line: 90\%).}
\label{SEvsGamma}
\end{figure}

We plot in Fig.  \ref{SEvsGamma} the photon index of the primary continuum as a function of the soft-excess strength $q$ (in the case of distant reflection). This plot reveals a hint of a weak correlation between $\Gamma$ and $q$. Indeed, Spearman rank analysis gives a correlation coefficient of r=0.19 and a null-hypothesis probability of about 10\%. A bootstrap on the data points gives a probability of 91\% of a positive correlation. We use the Bayesian \texttt{linmix\_err} procedure to perform a linear regression. We obtain
\begin{equation}
\Gamma=0.14\,_{-0.06}^{+0.05} \times q + 1.65\,_{-0.02}^{+0.04}
.\end{equation}
The intrinsic scatter of $\Gamma$ is 0.32. CI is the blue contour in Fig.  \ref{SEvsGamma}.

Removing the NLSy1s from our sample still gives a probability of having a positive correlation of 87\%. If we do not consider objects with warm absorber, we have a probability of 85\% of a positive correlation.

The possible faint correlation between the photon index of the primary power law and the soft-excess strength is also found in the case of reflection mainly owing to a local reflector with a Spearman correlation coefficient of r=0.14 and a null-hypothesis probability of 21\%. We find the relation:

\begin{equation}
\Gamma=0.11\,_{-0.03}^{+0.08} \times q + 1.71\,_{-0.06}^{+0.001}
.\end{equation}
The intrinsic scatter of $\Gamma$ is 0.30. 

If we only consider the ten NLSy1 objects of our sample, Spearman statistics give a correlation coefficient of r=0.50, a null-hypothesis probability of 14\%, and a probability of 96\% of a positive correlation.

This correlation between the photon index and the soft-excess strength is consistent with the result of \cite{Page2004}, who found a correlation with a  $\sim$ 4\% null-hypothesis probability obtained with seven objects observed by \textit{XMM-Newton}.

The red line in Fig.  \ref{SEvsGamma} represents the relation between $\Gamma$ and $q$ expected in the case of ionized reflection. Density contours of the \texttt{relxilllp\_ion} simulations described in Sect. \ref{4.2} are plotted in red (solid line: 68\%; dashed line: 90\%). Spearman statistics give a correlation coefficient of r=-0.07 and a null-hypothesis probability of $2\times10^{-5}$\%. The relation is

\begin{equation}
\Gamma=-0.07\,_{-0.01}^{+0.01} \times q + 1.99\,_{-0.008}^{+0.008}
.\end{equation}

\noindent The results of the simulations are inconsistent with our observational results.

\subsection{Reflection and photon index}
\label{5.2}

\begin{figure} [!b]
\resizebox{\hsize}{!}{\includegraphics[trim = 10mm 60mm 10mm 70mm, clip, angle=0]{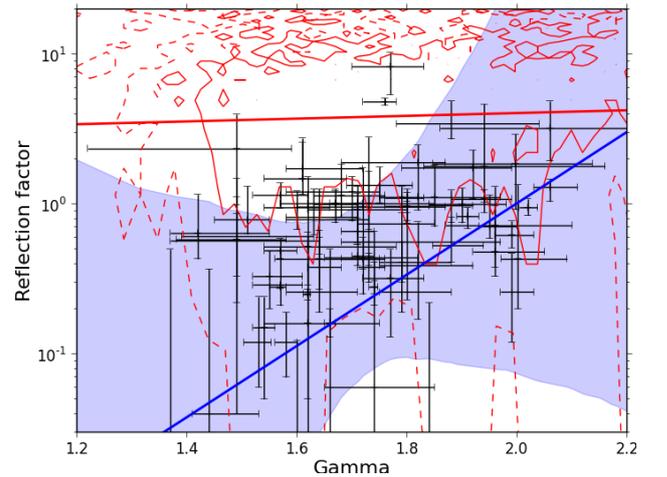}} 
\caption{Reflection factor against the power-law slope obtained with \textit{XMM-Newton} and \textit{BAT} data (in the case of distant reflection). CI is represented by the blue contour. The red line represents the relation between $R$ and $\Gamma$ in the case of ionized reflection. Density contours are plotted in red (solid line: 68\%, dashed line: 90\%).}
\label{RvsGamma}
\end{figure}

We plot in Fig. \ref{RvsGamma} the reflection factor as a function of the photon index (in the case of reflection mainly due to a distant reflector). Spearman statistics show that a positive correlation exists between $R$ and $\Gamma$ (r=0.35 with a null-hypothesis probability of 0.02\%). Using the Bayesian approach described in the previous sections, we perform a linear regression taking errors in x and y axes and intrinsic scatter into account. It results in the relation

\begin{equation}
R=1.07\,_{-0.37}^{+0.76}\times \Gamma -1.11\,_{-1.34}^{+0.62}
.\end{equation}

The intrinsic scatter is 0.17. 
The CI corresponding to this linear regression is represented by the blue contour in Fig. \ref{RvsGamma}. This correlation between reflection and photon index has already been observed in previous works \citep{Zdziarski1999,Lubinski2001,Perola2002,Mattson2007}.

When removing the NLSy1s from our sample, the positive correlation between $R$ and $\Gamma$ remains with a Spearman correlation coefficient of r=0.45 and a null-hypothesis probability of 0.006\%. Considering objects with soft excess without warm absorber, the probability of having a positive correlation is 99.2\%.

In the case of a local-dominated reflection, we also find a steeper relation between $\Gamma$ and $R$. Spearman statistics give a correlation factor of r=0.33 with a null-hypothesis probability of 0.2\%. The correlation is characterized by the equation

\begin{equation}
R=1.27\,_{-0.57}^{+0.23}\times \Gamma -0.93\,_{-0.27}^{+0.55}
.\end{equation}
The intrinsic scatter is 0.09. 

Considering only NLSy1s, we obtain a Spearman correlation coefficient of r=-0.11, a null-hypothesis probability of 76\% and a probability of 61\% of a negative correlation. 
However, the very flat slope of -0.007 obtained by linear regression is compatible, in view of its large uncertainties, with the positive relation found with the entire sample. The relation between $R$ and $\Gamma$ when we only consider that NLSy1s is not statistically significant.

The red line and contours in Fig. \ref{RvsGamma} are the results expected in the case of ionized reflection. We can see that, according to \texttt{relxilllp\_ion} simulations described in Sect. \ref{4.2}, a weak correlation is expected between $R$ and $\Gamma$ (Spearman correlation coefficient r=0.10 with a null-hypothesis probability of $1\times10^{-8}$\%). The relation is

 \begin{equation}
R=0.39\,_{-0.27}^{+0.27}\times \Gamma +5.03\,_{-0.55}^{+0.54}
.\end{equation}

\noindent The slope found between $R$ and $\Gamma$ in the case of blurred ionized reflection is different from the one found in the data.

\section{Spectra stacking}
Since we want to study the shape of the spectra for different soft-excess strengths in both soft and hard X-ray energy bands, we stack \textit{XMM-Newton EPIC/PN} and \textit{Swift/BAT} spectra for four groups divided on the basis of their soft-excess strengths, considering values of $q$ obtained when fitting the individual spectra in the case of distant-dominated reflection.  
The choice for the $q$ ranges is made in order to have an equivalent number of spectra in each group (see Table \ref{FitStack}).

\begin{figure} [!b]
\resizebox{\hsize}{!}{\includegraphics[trim = 30mm 30mm 40mm 40mm, clip, angle=0]{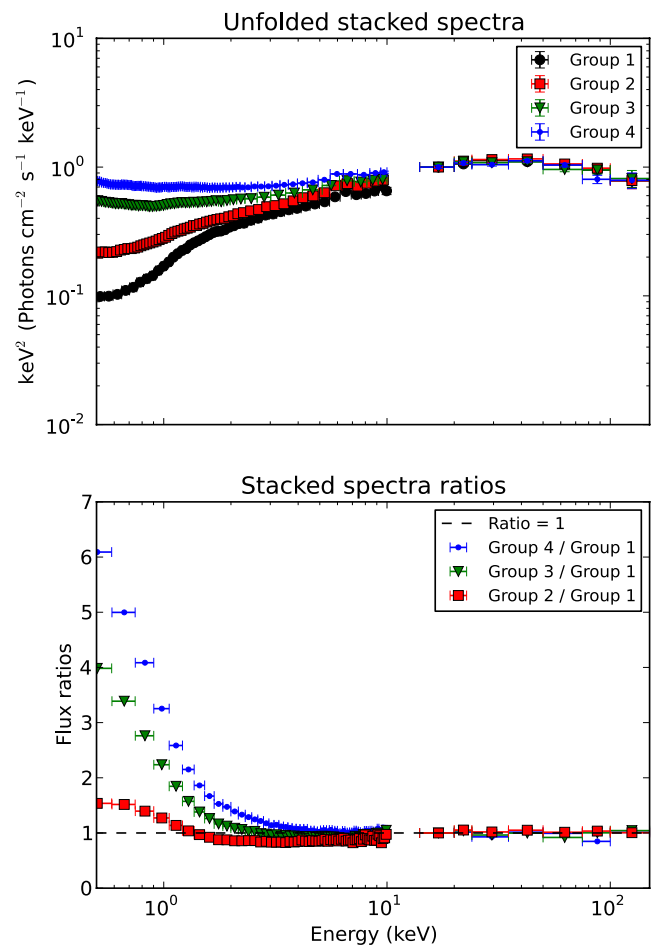}} 
\caption{Results of stacking spectra per soft-excess strengths. Top panel: Unfolded stacked \textit{XMM-Newton/PN} and \textit{Swift/BAT} spectra for the four groups of different $q$ (normalized to the spectrum with a lower $q$ value, at 15\,keV). The $q$ value is increasing for stacked spectra from black circles (Group 1) to red squares (Group 2), to green triangles (Group 3) and to blue dots (Group 4). Bottom panel: Ratios of the stacked \textit{XMM-Newton/PN} and \textit{Swift/BAT} spectra. The ratio of stacked spectra of higher $q$ (Group 4) over lower $q$ (Group 1) is represented in blue points. The ratio of stacked spectra from Group 3 over Group 1 is plotted in green triangles. The ratio of Group 2 over Group 1 is in red squares.}
\label{Stacking}
\end{figure}

Before stacking the spectra, we renormalize each \textit{PN} and \textit{BAT} spectra to the same flux,
 to avoid any preponderance of spectral shape from objects with higher fluxes. We then stack \textit{XMM-Newton/PN} and \textit{Swift/BAT} spectra per groups of soft-excess strengths, using the \texttt{addspec} tool from \texttt{FTOOLS} \footnote{\url{http://heasarc.gsfc.nasa.gov/ftools/}} \citep{Blackburn1995}. The top panel of Fig. \ref{Stacking} shows the resulting stacked \textit{PN} and \textit{BAT} spectra. 
The figure shows the evidence of different soft-excess strengths in the soft X-ray emission as expected, but no difference in the reflection strengths at higher energy.
We measure the photon index and the reflection factor for each of the stacked spectra by fitting them between 3 and 100\,keV with a \texttt{pexmon} model. The resulting parameters are presented in Table \ref{FitStack}. We see a clear increase in the value of the photon index per increasing soft-excess strength (as the possible correlation shown in Sect. \ref{5.1}), but we do not see any trend for the reflection factor, showing that the reflection strength and the soft-excess strength are not linked. The photon indexes vary when fitting the \textit{XMM-Newton/PN} and \textit{Swift/BAT} stacked spectra of the four groups of different soft-excess strengths together, although \textit{BAT} spectra all look the same, because \textit{XMM-Newton/PN} spectra have a more important weight in the fit than \textit{BAT} spectra (because of their better signal-to-noise ratios). The value of $\Gamma$ then strongly depends on the shape of \textit{XMM-Newton} spectra, which look different between 3 and 10 keV for the different soft-excess strengths. When fitting \textit{BAT}-stacked spectra alone, we cannot see any trend for $\Gamma$ and $R$ as a function of $q$, as photon indexes are similar for the four groups, and reflection factors have failed to be constrained. The absence of evolution of $R$ in the stacked spectra, which is in contradiction with the anti-correlation found in our sample between $R$ and $q$ (see Sect. \ref{4.1}), may be due to low statistics.

\begin{table*}
\caption{Results of the fitting on the stacked spectra using, on one hand, all the objects of our sample with a soft excess, and on the other hand, NLSy1s alone.}\label{FitStack}
\begin{center}
\begin{tabular}{|cc|ccc|ccc|}
\cline{3-8}
\multicolumn{2}{c|}{} & \multicolumn{3}{c|}{Total sample} & \multicolumn{3}{c|}{NLSy1s} \\
\hline
Group & Soft-excess strength & Objects & $R$ & $\Gamma$ &Objects & $R$ & $\Gamma$ \\
\hline
1 &$q<0.25$& 19 & 0.69 $\pm$ 0.16 & 1.75 $\pm$ 0.05& 2 & 0.53 $\pm$ 0.27 & 1.79 $\pm$ 0.06\\
2 & $0.25\leq q<0.4$& 19 & 0.97 $\pm$ 0.23 & 1.81 $\pm$ 0.06& 3 & 1.01 $\pm$ 0.71 & 1.98 $\pm$ 0.08\\
3 & $0.4\leq q<0.6$ &20 & 0.60 $\pm$ 0.29 & 1.99 $\pm$ 0.08 &2 & 0.59 $\pm$ 0.28 & 2.15 $\pm$ 0.08\\
4 & $q>0.6$ &21 & 0.84 $\pm$ 0.16 & 2.04 $\pm$ 0.04& 3 & 0.47 $\pm$ 0.10 & 2.29 $\pm$ 0.03\\
\hline
\end{tabular}
\end{center}
\end{table*}

The bottom panel of Fig. \ref{Stacking} shows the ratios of stacked spectra, calculated with \texttt{mathpha} \citep{Blackburn1995}. In both panels,  
we can see that the difference in spectral shape in the soft X-ray band is obvious for different soft-excess strengths. However, we do not notice any spectral shape difference in the hard X-ray band. 

If we consider only the NLSy1s present in our sample and stack their \textit{XMM-Newton} and \textit{Swift/BAT} spectra for four groups of soft-excess strength, we observe the same trend as the one found for the entire sample. We see in Table \ref{FitStack} that the photon index is increasing for an increasing value of the soft-excess strength, but we do not note any evolution trend for the reflection factor.


\section{Relation with the Eddington ratio}
For 20 objects of our sample, Eddington ratios have been calculated by \cite{Ricci2013}. They used average bolometric corrections $k_{x}$ (taken from the literature) obtained from studies of the AGN spectral energy distribution to calculate the Eddington ratios. Considering black-hole masses from previous works of \cite{Woo2002} and \cite{Vasudevan2010}, we can easily calculate the Eddington ratios for 13 additional objects not studied in \cite{Ricci2013}. 
The Eddington ratio is calculated as

\begin{equation}
\lambda_{Edd}=\frac{L_{Bol}}{L_{Edd}} = \frac{k_{x}\times L_{2-10\,keV}}{1.26\times 10^{38} \times M_{BH}/M_{\odot}}
.\end{equation}

\begin{figure} [!b]
\resizebox{\hsize}{!}{\includegraphics[trim = 10mm 60mm 10mm 70mm, clip, angle=0]{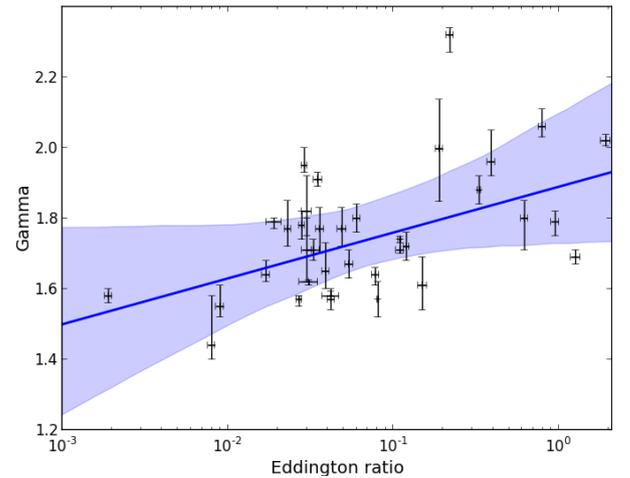}} 
\caption{Photon index of the primary continuum as a function of the Eddington ratio. Linear regression is represented by the blue line and 99\% CI is given by the blue contour.}
\label{GvsLambda}
\end{figure}

\noindent We plotted the photon index as a function of the Eddington ratio for the 33 NLSy1, Sy1, and Sy1.5 objects in Fig. \ref{GvsLambda}. A linear regression with \texttt{linmix\_err} 
gives the following relation:

\begin{equation}
\Gamma=0.13\,_{-0.04}^{+0.05}\times log(\lambda_{Edd}) +1.89\,_{-0.03}^{+0.08}
\end{equation}

\noindent with a correlation coefficient of 0.47 and a null-hypothesis probability of 0.4\%. The instrinsic scatter of $\Gamma$ is 0.02.
Such a positive correlation has already been found in several works performed using \textit{ASCA}, \textit{ROSAT}, \textit{Swift}, and \textit{XMM-Newton} \citep{Wang2004,Grupe2004,Porquet2004,Bian2005,Grupe2010}, establishing a relation between $\Gamma$ and $\lambda_{Edd}$ \citep{Shemmer2008,Risaliti2009,Jin2012}.

\begin{figure} [!h]
\resizebox{\hsize}{!}{\includegraphics[trim = 10mm 60mm 10mm 70mm, clip, angle=0]{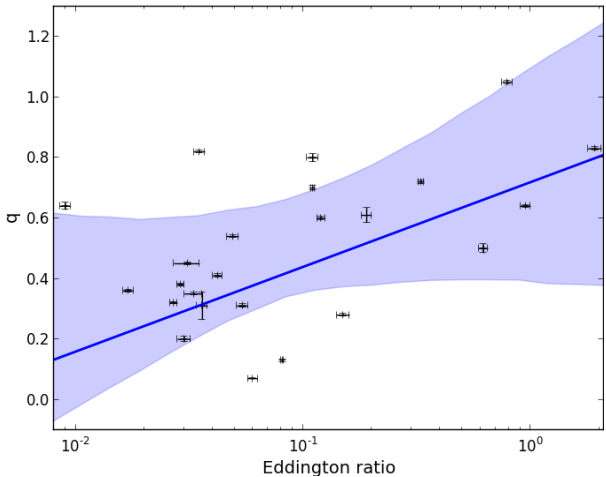}} 
\caption{Soft-excess strength as a function of the Eddington ratio. Linear regression is represented by the blue line and 99\% CI is given by the blue contour.}
\label{QvsLambda}
\end{figure}

In Fig. \ref{QvsLambda} we plotted the soft-excess strength as a function of the Eddington ratio. We find
\begin{equation}
q=0.28\,_{-0.15}^{+0.01}\times log(\lambda_{Edd}) +0.72\,_{-0.09}^{+0.11}
.\end{equation}
A correlation exists with a coefficient of 0.43 and is significant (null-hypothesis probability of 1\%). The intrinsic scatter of $q$ is 0.07.

\begin{figure} [!h]
\resizebox{\hsize}{!}{\includegraphics[trim = 10mm 60mm 10mm 70mm, clip, angle=0]{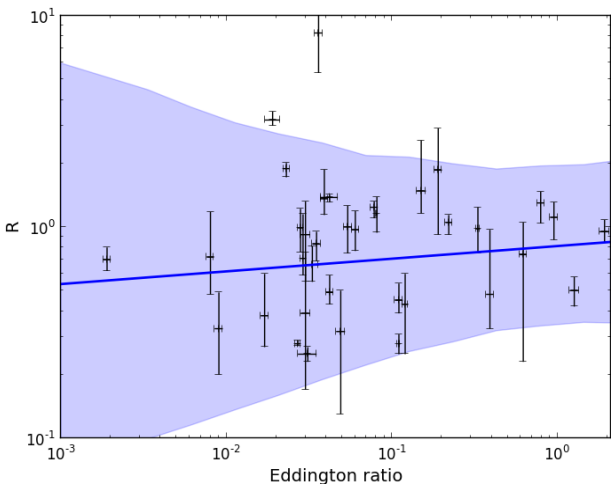}} 
\caption{Reflection factor as a function of the Eddington ratio. Linear regression is represented by the blue line and 99\% CI is given by the blue contour.}
\label{RvsLambda}
\end{figure}

We plotted the reflection factor as a function of the Eddington ratio in Fig. \ref{RvsLambda}. The linear regression process leads to the relation:
\begin{equation}
log(R)=0.06\,_{-0.08}^{+0.09}\times log(\lambda_{Edd}) -0.09\,_{-0.12}^{+0.10}
.\end{equation}
The correlation between $R$ and $\lambda_{Edd}$ is not significant since the correlation coefficient of 0.17 is obtained with a null-hypothesis probability of 31\%. The intrinsic scatter of $log(R)$ is 0.05.

\section{Discussion}
\label{9}

We studied 102 Seyfert 1s (Sy1.0, 1.2, 1.5, NLSy1) from the \textit{Swift}/\textit{BAT} 70-Months Hard X-ray Survey catalog, using \textit{Swift}/\textit{BAT} and \textit{XMM-Newton} observations. The simultaneous spectral analysis of the soft and the hard X-ray emission aimed to study the behavior between the soft excess present in a large number of Seyfert 1s and the reflection component. We have seen that our results are not affected by the presence of NLSy1s in our sample, because the trends are similar or compatible when excluding these objects from our analysis or when considering only NLSy1s (relations found between $q$, $R$, and $\Gamma$ and stacking of spectra). This suggests that the mechanism responsible for the soft excess is similar for all the categories of our sample. We did not find any effect of ionized absorption present in many objects of our sample. 
Our method of measuring the soft-excess strength is  robust  even in objects with moderate absorption.

\subsection{Link between soft excess and reflection}
\label{9.1}

\cite{Vasudevan2013} studied a sample of AGNs from the 58-month \textit{Swift}/\textit{BAT} catalog. Using \textit{BAT} and \textit{XMM-Newton} data, they constrained the reflection and the soft-excess strength in 39 sources. A plot of the reflection strength $R$ against the soft-excess strength for the 23 low-absorption sources from \cite{Vasudevan2013} is presented in Figure 1 of \cite{Vasudevan2014}. The figure shows strong hints of a correlation. This correlation can be explained by the fact that, in the case of local reflection,  higher reflection leads to stronger emission lines below 1\,keV that are smeared in the vicinity of the supermassive black hole, inducing a larger smooth bump at low energy, the soft excess.

It is necessary to use simultaneous broad-band data in order to distinguish between mechanisms at the origin of the soft excess. In \cite{Vasudevan2014}, \textit{XMM-Newton} and \textit{NuSTAR} simulations are done to produce a plot of the strength of the hard X-ray emission (measured by a neutral reflection model) versus the strength of the soft excess (modeled by a blackbody). This figure can be used as a diagnostic plot to determine the soft-excess production mechanism. Indeed, there is no evidence of a correlation between $R$ and the soft-excess strength in the case of ionized absorption, for example. But a correlation exists in the case of ionized reflection. 

In our work, we use a larger sample than \cite{Vasudevan2014} (our sample contains $\sim$ 3.5 times more sources) to produce a plot of the reflection factor versus the soft-excess strength, but we did not obtain the same results as the similar plot of \cite{Vasudevan2014}. Our work has ten objects in common with their analysis \citep{Vasudevan2013}. For four of them (NGC 4593, NGC 5548, IC 2637, and Mrk 50), $R$ and $q$ values match both studies. For four other sources (Mrk 817, KUG 1141+371, QSO B1419+480, and NGC 4235), very low $R$ and/or $q$ values from the study of \cite{Vasudevan2014} have intermediate $R$ and $q$ values in our work. For NGC 4051, we obtain a smaller reflection factor than \cite{Vasudevan2014}, and non-negligible absorption prevents the detection of a soft excess. Mrk 766 gives us a similar value for the soft-excess strength, but a lower value of $R$, compared to results from \cite{Vasudevan2014}. We note that, in general, reflection factors measured in our work are rarely higher than 2, as spectral fitting results from \cite{Crummy2006} and \cite{Walton2013}, who obtained $R\sim1$ on average, while \cite{Vasudevan2013} get higher $R$ values. The correlation found in \cite{Vasudevan2014} seems to be driven by objects that show extreme values of $R$ and $q$ but whose measurements are unreliable.
These differences between the results can occur because \cite{Vasudevan2013} used a different fitting process. Indeed, they fitted spectra from 1.5\,keV with a \texttt{pexrav} model, ignoring the iron energy band (between 5.5 and 7.5 keV), fixing abundances, adding an absorber to the model and using a blackbody to model the soft excess from 0.4\,keV. They also used a different definition of $q$ ($L_{BB, 0.4-3\,keV}/L_{pow, 1.5-6\,keV}$). The  cross-calibration method also differs. Indeed, \cite{Vasudevan2013} renormalized some \textit{BAT} spectra for objects whose soft X-ray observations had been taken within the timeframe of the \textit{BAT} survey, using the \textit{BAT} light curve and considering the ratio between \textit{BAT} flux during the entire survey and \textit{BAT} flux during the \textit{XMM-Newton} observation. However, variability in AGN can happen also on time scales shorter than a month, so this ratio might not be indicative of the real difference in flux between the short \textit{XMM-Newton} observations and the 70-month averaged \textit{Swift/BAT} flux.

After analyzing our sample of objects spectrally, we do not see any hint of correlation as found in the plot of the reflection versus the soft-excess strength of \cite{Vasudevan2014}. We even find evidence of a weak anti-correlation between $R$ and $q$ (see blue contours in top panel of Fig. \ref{RvsSE}). We used the \texttt{relxilllp\_ion} model to simulate spectra, as we measure $q$ and $R$ differently than in the work of \cite{Vasudevan2014}. Our simulations are in good agreement with those carried out by \cite{Vasudevan2014}. Indeed, as reported by \cite{Vasudevan2014}, if the soft excess was entirely due to blurred reflection, one would expect to find a correlation between the reflection factor and the soft-excess strength (see red line in top panel of Fig. \ref{RvsSE}). In this case, we should obtain high $R$ and high $q$, as well as small $R$ and small $q$ ($R<2.0$ and $q<0.75$, blue box in bottom panel of Fig. \ref{RvsSE}). That most of the objects of our sample (75\%) appear in the blue box is difficult to explain in the blurred ionized-reflection hypothesis, because this model predicts that only 25\% of the objects should appear as small-$q$, small-$R$. There is no easy way to restrict the parameter space to match the blue box. In particular, high-$R$, high-$q$ objects are objects with intermediate ionization. 
Considering our sample of 79 Seyfert 1s, no positive correlation is found between $R$ and $q$ (both in the local and in the distant reflection scenarii) and only low $R$ and $q$ values are measured (see bottom panel of Fig. \ref{RvsSE}). These contradictions are a strong argument against the ionized-reflection hypothesis as the origin of the soft excess in most objects.

By stacking \textit{XMM-Newton/PN} and \textit{Swift/BAT} spectra in different bins of soft-excess intensity (see Fig. \ref{Stacking}), we have seen that there is no obvious difference in the spectral shape in the hard X-ray band in the individual stacked spectra and in their ratios. 
In stacked spectra with the lower soft-excess strength values, absorption can be seen below 1\,keV (see black circles and red squares in Fig. \ref{Stacking}), as many objects of our sample are slightly absorbed. For higher values of the soft-excess intensity (see green triangles and blue points of Fig. \ref{Stacking}), the soft-excess emission compensates for and dominates the absorption, assuming that absorption is equally present in each of the four groups of soft-excess strength.
While the spectra are steeper for a stronger soft excess, we do not see any evolution of the reflection factor. We observe the same results when only stacking NLSy1 spectra.
The values of the photon index we obtain are consistent with those found by \cite{Ricci2011} by stacking \textit{INTEGRAL} spectra of type 1 and 1.5 AGN.
There is no stronger reflection feature in the hard X-ray band for stronger soft excess, since we would expect in the case where ionized reflection is at the origin of the soft excess. This result suggests that the blurred ionized reflection is not responsible for the existence of the soft excess in most objects of our sample.

\subsection{Soft excess and X-ray continuum}

Studying our 79 objects with soft excess, we found a possible correlation between the photon index of the primary continuum and the strength $q$ of the soft excess (see blue contours in Fig. \ref{SEvsGamma}). This trend is also verified when we consider the objects grouped per soft-excess strength, by fitting spectra stacked per soft-excess strengths (see Table \ref{FitStack}). Such a correlation  has already been cautiously presented in \cite{Page2004} with a higher significance than in this work. 
A possible faint anti-correlation between $\Gamma$ and $q$ is expected in the case of ionized reflection, as shown by \texttt{relxilllp\_ion} simulations (see red line and contours in Fig. \ref{SEvsGamma}), which is at odds with the possible positive correlation found with our data.
A positive correlation like the one found in our sample might suggest a connection between the soft excess and the cooling of the hot corona. As proposed in the case of Mrk 509 \citep{Pop2013}, a warm corona could upscatter the optical-UV photons from the accretion disk to produce the soft excess. This soft excess could constitute a non-negligible part of the soft photons Comptonized in the hot corona (producing the primary X-ray continuum), since they would participate in its cooling. Indeed, the warm plasma emission peaks from a few eV to a few hundred eV, so it could be considered by the hot corona as a soft photon field with an intermediate temperature (see Figure 10 in \citealt{Pop2013}). The geometry of a hot photon-starved corona surrounded by an outer cold disk has already been suggested by \cite{Abramowicz1995} and \cite{Narayan1995} and applied, for example, in NGC 4151 by \cite{Lubinski2010}. \cite{Pop2013} propose that the plasma responsible for the soft excess in Mrk 509 could be a warm upper layer of this accretion disk. In this case, a higher soft-excess strength $q$ means a more efficient cooling and a softer X-ray emission. The relation between $\Gamma$ and $q$ is an argument in favor of warm Comptonization models to explain the soft excess.

Studying the relation between $R$ and $\Gamma$, we found a correlation with our 79 objects (see blue contours in Fig. \ref{RvsGamma}). A faint correlation is also expected with ionized reflection (see red contours in Fig. \ref{RvsGamma}), but the relation slopes are different. The mean value of $R\sim5$ for the ionized reflection is due to the chosen parameters for the simulations. Indeed, we chose a reflection factor going up to 100 for the \texttt{relxilllp\_ion} model, so the reflection factor R measured with \texttt{pexmon} is high. But this mean value of $R\sim5$ resulting from simulations does not have a real meaning because the values of the other parameters are not known.
Correlations have been found previously between the reflection and the power-law photon index in several works. 
Using \textit{Ginga} spectra of radio-quiet Seyfert 1s and narrow emission-line galaxies, \cite{Zdziarski1999} found a very strong correlation between the intrinsic spectral slope in X-rays and the amount of Compton reflection from a cold medium ($R\propto \Gamma^{12}$). They interpreted this as due to a feedback within the source where the cold medium responsible for the reflection emits soft photons that irradiate the X-ray source and participate in the cooling as seeds for Compton upscattering. Fainter correlations have
also been found between $R$ and $\Gamma$ and between the spectral slope and the strength of the iron line \citep{Lubinski2001,Perola2002}. \cite{Mattson2007} found a relation between $R$ and $\Gamma$ ($R=0.54\Gamma-0.87$), close to the one we found during our analysis ($R=1.07\,_{-0.37}^{+0.76}\times \Gamma -1.11\,_{-1.34}^{+0.62}$), which could be due to degeneracies during modeling process. According to \cite{Malzac2002}, the correlation between $R$ and $\Gamma$ could be due to the presence of a remote cold material. Indeed, assuming that disk reflection is negligible and thus that reflection is mainly due to distant cold material, fluctuations in the primary X-ray emission slope at constant flux make the spectrum pivot, inducing a correlation between $\Gamma$ and $R$. 
The correlation between $R$ and $\Gamma$ that we find for both local and distant reflection scenarii, because it differs from blurred ionized-reflection model expectations in slope, is another argument against the ionized-reflection hypothesis as the origin of the soft excess. 

\subsection{Relations with the Eddington ratio $\lambda_{Edd}$}
A positive correlation has already been found between the photon index of the primary continuum and the Eddington ratio in several works \citep{Wang2004,Grupe2004,Porquet2004,Bian2005,Kelly2007,Shemmer2008,Risaliti2009,Grupe2010,Jin2012}. \cite{Shemmer2008} found the relation $\Gamma=0.31 \,log(\lambda_{Edd})\,+\,2.11$, consistent with results from \cite{Wang2004} and \cite{Kelly2007}. \cite{Risaliti2009} and \cite{Jin2012} find a steeper relation: $\Gamma \propto 0.60 \,log(\lambda_{Edd})$.  This relation implies a link between the accretion disk and the hot corona that could be due to the fact that the more accretion disk emits optical/UV photons, the more efficient the cooling of the corona. 
With our sample of 79 Seyfert 1s, we possibly found such a correlation between $\Gamma$ and $\lambda_{Edd}$ (see Fig. \ref{GvsLambda}), with a flatter relation that is inconsistent with previous results ($\Gamma=0.13\, log(\lambda_{Edd}) +1.89$). The difference between our result and previous ones may be due to the different fitting procedures, because \cite{Shemmer2008}, \cite{Risaliti2009} and \cite{Jin2012} used a simple absorbed power law to fit their data.

As shown in Fig. \ref{QvsLambda}, we found a possible correlation between  $\lambda_{Edd}$ and $q$. This might show a link between the accretion disk and the warm corona responsible for the soft excess in the warm Comptonization model. In the case of Mrk 509, \cite{Pop2013} propose a geometry where the warm corona is the top layer of the accretion disk. The warm corona heats the deeper layers and Comptonizes their optical/UV photons, creating the soft-excess feature (see Figure 10 in \citealt{Pop2013}). Such a geometry could then explain the correlation between $\lambda_{Edd}$ and $q$, which is an argument in favor of the warm Comptonization hypothesis as the origin of the soft excess. Since we find correlations between $\lambda_{Edd}$ and $q$ and between  $\Gamma$ and $\lambda_{Edd}$, a correlation between $\Gamma$ and $q$, such as the one found in Sect. \ref{5.1}, is expected, driven by $\lambda_{Edd}$.  

The Baldwin effect, which is an anti-correlation between the equivalent width (EW) of the iron K$\alpha$ line and the X-ray luminosity ($EW \propto L_{X}^{-0.20}$ -- \citealt{Iwasawa1993}), could be explained by the decrease in luminosity when the covering factor of the torus from the unification model increases \citep{Page2004b,Bianchi2007}. A similar trend has been observed between the EW of the FeK$\alpha$ line and the Eddington ratio. Studying a large sample of unabsorbed AGN, \cite{Bianchi2007} found the relation $EW \propto \lambda_{Edd}^{-0.19}$, while \cite{Shu2010} used a sample of \textit{Chandra/HEG} observations to find a similar relation of $EW \propto (L_{2-10\,keV}/L_{Edd})^{-0.20}$ when fitting each observation, but a weaker correlation ($EW \propto (L_{2-10\,keV}/L_{Edd})^{-0.11}$) when doing the fits per source. Using results of \cite{Shu2010} and bolometric corrections, \cite{Ricci2013} found the relation $log(EW) = -0.13 \, log(\lambda_{Edd})+1.47$.

We found, in our work, an anti-correlation between $R$ and $q$. We also found a positive correlation between $q$ and $\lambda_{Edd}$. We then expect to have an anti-correlation between $R$ and $\lambda_{Edd}$, which is similar to the Baldwin effect, since the EW of the Fe K$\alpha$ line and the reflection factor $R$ are both representative of the reflection strength. Unfortunately, this expected anti-correlation between $R$ and $\lambda_{Edd}$ is not seen in our sample,
probably because of intrinsic scatter and large uncertainties. 

A possible explanation for this anti-correlation between $R$ and $q$ is the warm Comptonization. In this scenario, a warm plasma, which could be the upper layer of the disk, upscatters the soft optical/UV photons from the disk to reproduce the soft excess. 
Since the warm plasma at a temperature of $\sim$ 1\,keV is highly ionized, reflection on this medium is largely featureless and follows the primary emission. Therefore, a disk covered with a warm plasma sees its reflection factor $R$ decrease compared to the case with little warm plasma or none at all. As a result, the stronger the soft excess, the smaller the reflection factor. 
The anti-correlation found in data between $R$ and $q$ could then be explained if the soft excess came from a warm plasma.

\section{Conclusion}
\label{7}

The nature of the soft excess in AGN is still uncertain because physical mechanisms used to model this feature are difficult to distinguish when analyzing soft X-rays spectra. The \textit{Swift}/\textit{BAT} and \textit{XMM-Newton} spectral analysis of a large sample of Seyfert 1s from the \textit{Swift}/\textit{BAT} 70-Months Hard X-ray Survey catalog allowed a hard X-ray view of the soft excess in AGN. We fit the 3-100 keV data with phenomenological model and then quantify the soft excess below 2 keV with respect to this model. We found that 80\% of the objects of our sample show the presence of a soft excess. 

Fitting the spectra of 79 Seyfert 1s lowly absorbed and showing a soft excess, we showed  that the soft-excess strength and the reflection factor are not positively correlated. By stacking \textit{XMM-Newton/PN} and \textit{Swift/BAT} spectra per soft-excess strengths, we have shown that the reflection characterized by the Compton hump at about 30\,keV does not vary with the soft-excess strength.
These results contradict the correlation expected from ionized reflection, as shown by our simulations with \texttt{relxilllp\_ion} and by simulations of \textit{XMM-Newton} and \textit{NuSTAR} spectra from \cite{Vasudevan2014}. 
This contradiction between expectations and measurements is a strong argument against the ionized-reflection hypothesis as the origin of the soft excess in most objects.
The possible anti-correlation we found could be explained by a warm Comptonization scenario, where a warm plasma covering the disk would make the reflection featureless.

We have also seen that the strength of the soft excess $q$ is correlated with the spectral index $\Gamma$ and with the Eddington ratio $\lambda_{Edd}$. This could be explained by warm Comptonization scenarii, such as the one described in \cite{Pop2013}, where a higher $q$ value might mean a more efficient cooling of the hot corona responsible for the primary X-ray emission and hence a steeper spectrum. Furthermore, the relation found between $R$ and $\Gamma$ is different from the one found in \texttt{relxilllp\_ion} simulations and can be used as an additional argument against ionized reflection. The correlation could be because the medium responsible for reflection emits soft photons that participate in the cooling of the hot corona.

The relation found between $R$, $\Gamma$, and $q$ are found when assuming a reflection component mainly due to distant material, as well as if this reflection mainly comes from the accretion disk, which we cannot distinguish here.

This work suggests that the soft excess present in 80\% of the objects of our sample is, in most cases, likely not due to blurred ionized reflection, but can most probably be explained by warm Comptonization. Future works with \textit{NuSTAR} and \textit{ASTRO-H} will shed light on this issue, as the better signal-to-noise data they will provide in the hard X-ray band may allow both models to be spectrally distinguished.

\appendix


\longtab{2}{
\begin{center}
\begin{longtable}{lccccccc}

\caption{List of the sources used for this study, with their spectral types, redshifts $z$, Galactic column densities ($N_{\rm\,H}^{\rm\,G}$ values from \citet{Dickey:1990fj}), and soft X-ray observations information (observation date, observation identification and net exposure).}\label{tab:epic_info}\\
\hline \hline \\

\noalign{\smallskip}
   \multicolumn{1}{l}{Source} &
\multicolumn{1}{c}{Type} &
\multicolumn{1}{c}{z} &
\multicolumn{1}{c}{$N_{\rm\,H}^{\rm\,G}$} &
   \multicolumn{1}{c}{Obs. date} &
   \multicolumn{1}{c}{Obs. ID} &
   \multicolumn{1}{c}{Net exposure} \\

\noalign{\smallskip}
   \multicolumn{1}{l}{ } &
   \multicolumn{1}{c}{} &
\multicolumn{1}{c}{ } &
\multicolumn{1}{c}{[{\tiny $10^{20} \rm \,cm^{-2}$}] } &
\multicolumn{1}{c}{ \tiny{YYYY-MM-DD} } &
\multicolumn{1}{c}{ } &
   \multicolumn{1}{c}{[\tiny{$\rm\,ks$}]} \\
\noalign{\smallskip}

\hline
\noalign{\smallskip}

\endfirsthead

\multicolumn{5}{l}%
{\small \hspace{-0.1in}{{\textbf{\tablename\ \thetable{}.} Information on sources and observations used. -- \textit{continued} }}} \\
\noalign{\smallskip}
\hline \hline \\[0.05ex]

\noalign{\smallskip}
   \multicolumn{1}{l}{Source} &
\multicolumn{1}{c}{Type} &
\multicolumn{1}{c}{z} &
\multicolumn{1}{c}{$N_{\rm\,H}^{\rm\,G}$} &
   \multicolumn{1}{c}{Obs. date} &
   \multicolumn{1}{c}{Obs. ID} &
   \multicolumn{1}{c}{Net exposure} \\

\noalign{\smallskip}
   \multicolumn{1}{l}{ } &
   \multicolumn{1}{c}{} &
\multicolumn{1}{c}{ } &
\multicolumn{1}{c}{[{\tiny $10^{20} \rm \,cm^{-2}$}] } &
\multicolumn{1}{c}{ \tiny{YYYY-MM-DD} } &
\multicolumn{1}{c}{ } &
   \multicolumn{1}{c}{[\tiny{$\rm\,ks$}]} \\
\noalign{\smallskip}

\hline 
\noalign{\smallskip}

\endhead

\noalign{\smallskip}
\hline
\endfoot

\endlastfoot

1H 0419$-$577 & Sy 1.5 & 0.104 & 1.83 &2010$-$05$-$30 & 0604720301 & 100.3 \\
1H 2251$-$179$^P$  & Sy 1.5 & 0.064 & 2.7 & 2002$-$05$-$18 & 0012940101  & 61.8  \\
1RXS J213944.3+595016  &  Sy 1.5 & 0.114 & 59.7 &2008$-$05$-$11& 0555321001 & 8.7 \\
2MASSi J1031543$-$141651 &  Sy 1.0 & 0.086 & 6.45 &2004$-$12$-$19 & 0203770101 & 34.6 \\
2MASX J18560128+1538059 &  Sy 1.0 & 0.084 & 37.4 &2009$-$04$-$06 & 0550451601 & 6.2 \\
2MASX J22484165$-$5109338 &  Sy 1.5 & 0.100 & 1.35 &2007$-$05$-$15 & 0510380101 & 64.5 \\
3C 111.0 &  Sy 1.0 & 0.048 & 32.2 &2009$-$02$-$15 & 0552180101 & 71.8 \\
3C 382 &  Sy 1.0 & 0.058 & 7.46 &2008$-$04$-$28 & 0506120101 & 32.4 \\
3C 390.3 & Sy 1.5 & 0.056 & 4.28 & 2004$-$10$-$08 & 0203720201 & 51.7 \\
4C +74.26 &  Sy 1.0 & 0.104 & 12.2 &2004$-$02$-$06 & 0200910201 & 31.9 \\
4U 0517+17  & Sy 1.5 & 0.018 & 22.0 & 2007$-$08$-$21 & 0502090501 & 57.2 \\
6dF J2132022$-$334254$^P$ & Sy 1.2 & 0.030 & 4.07 & 2004$-$10$-$30 & 0201130301 & 46.0 \\
Ark 120 & Sy 1.0 & 0.032 & 12.6 & 2003$-$08$-$25 &  0147190101 & 105.3  \\
CGCG 229$-$015 & Sy 1.0 & 0.028 & 6.25& 2011$-$06$-$05 & 0672530301 & 23.9 \\
ESO 140$-$43 & Sy 1.5 & 0.014 & 7.3& 2005$-$09$-$08  &  0300240401  & 21.8  \\
ESO 141$-$55$^P$ & Sy 1.2 & 0.037 & 5.1 & 2007$-$10$-$30 &  0503750101 & 77.6  \\
ESO 198$-$024 & Sy 1.0 & 0.046 & 3.05 & 2006$-$02$-$04 & 0305370101 & 121.9 \\
ESO 209$-$12 & Sy 1.5 & 0.040 & 23.8 & 2006$-$03$-$25 &  0401790301 &  7.2   \\
ESO 323$-$77 &   Sy 1.2 & 0.015 & 7.4 &  2006$-$02$-$07 &  0300240501  &  25.6  \\
ESO 359$-$ G 019 & Sy 1.0 & 0.055 & 1.02 & 2004$-$03$-$09 & 0201130101 & 24.0 \\
ESO 548$-$G081$^P$ & Sy 1.0 & 0.014 & 3.04 & 2006$-$01$-$28 & 0312190601 & 10.0 \\
EXO 055620$-$3820.2 &  Sy 1.2 & 0.034 & 4.0 &  2006$-$11$-$03 &  0404260301  &  75.9  \\
Fairall 1116 & Sy 1.0 & 0.058 & 3.09 & 2005$-$08$-$28 & 0301450301 & 20.1 \\
Fairall 1146 & Sy 1.0 & 0.032 & 40.3 &  2006$-$12$-$12 &  0401790401  &  11.6  \\
Fairall 9 & Sy 1.2 & 0.047 & 3.28 & 2009$-$12$-$09 & 0605800401 & 129.6 \\
GQ Com & Sy 1.2 & 0.165 & 1.67 & 2002$-$05$-$30 & 0109080101 & 13.3 \\
GRS 1734$-$292$^P$& Sy 1.0 & 0.021 & 76.7 & 2009$-$02$-$26 &  0550451501  &  12.1  \\
$[$HB89$]$ 0052+251 & Sy 1.2 & 0.154 & 4.93 & 2005$-$06$-$26 & 0301450401 & 19.8 \\
$[$HB89$]$ 0119$-$286 &Sy 1.0 & 0.116 & 1.65& 2003$-$01$-$07 & 0110950201 & 5.7 \\
$[$HB89$]$ 0241+622 &Sy 1.2 & 0.044 & 74.2 &  2008$-$02$-$28 & 0503690101 & 30.0 \\
IC 0486 & Sy 1.0 & 0.027 & 3.95& 2007$-$10$-$28 & 0504101201 & 20.1 \\
IC 2637 & Sy 1.5 & 0.029 & 2.66& 2009$-$12$-$20 & 0601780201 & 13.2 \\
IC 4329A & Sy 1.2 & 0.016 & 4.4 &  2003$-$08$-$06 &  0147440101  &  118.4  \\
IGR J00335+6126  & Sy 1.5 & 0.105 & 61.6&  2010$-$01$-$15 & 0601740101    &  21.5  \\
IGR J07597$-$3842  & Sy 1.2 & 0.040 & 60.3 &  2006$-$04$-$08  & 0303230101    & 15.0   \\
IGR J11457$-$1827   & Sy 1.5 & 0.033 & 3.5 & 2004$-$06$-$08  & 0201130201   &  31.0  \\
IGR J12172+0710 & Sy 1.2 & 0.008 & 1.5 &  2004$-$06$-$09  & 0204650201   & 9.3   \\
IGR J13038+5348$^P$ & Sy 1.2 & 0.029 & 1.6 & 2006$-$06$-$23  & 0312192001   &  9.6  \\
IGR J13109$-$5552 & Sy 1.0 & 0.104 & 27.6 & 2009$-$02$-$26  & 0550450901  &  17.9  \\
IGR J16119$-$6036 &  Sy 1.5 & 0.016 & 23.1 &2009$-$02$-$18  & 0550451101  & 13.1   \\
IGR J16185$-$5928	&	NLSy 1 & 0.035 & 24.7 &	2009$-$02$-$18	&	0550451201	&	17.4	\\
IGR J16482$-$3036 & Sy 1.0 & 0.031 & 17.6 & 2006$-$03$-$01 & 0305831001  & 7.5   \\
IGR J16558$-$5203 & Sy 1.2 & 0.054 & 30.4 & 2006$-$03$-$01 & 0306171201  & 8.9   \\
IGR J17418$-$1212	& Sy 1.2 & 0.037 & 20.9 &		2006$-$04$-$04		&	0303230501	&	13.1			\\
IGR J17488$-$3253	&Sy 1.0 & 0.020 & 53.0 &		2007$-$03$-$03		&	0405390101	&	6.57			\\
IGR J18027$-$1455	& Sy 1.0 & 0.035 & 49.7 &		2006$-$03$-$25		&	0303230601	&	18.1			\\
IGR J18259$-$0706	& Sy 1.0 & 0.037 & 71.2 &		2011$-$03$-$07		&	0650591501	&	25.9			\\
IGR J19378$-$0617	& Sy 1.5 & 0.011 &	14.8 &	2009$-$04$-$28	&	0550451701	&	17.4	\\
IGR J21277+5656	& NLSy 1 & 0.015 & 78.7 &		2010$-$11$-$29	&	0655450101	&	127.5	\\
IRAS 04392$-$2713 & Sy 1.5 & 0.084 & 2.49 & 2005$-$08$-$13 & 0301450101 & 20.0 \\
IRAS 15091$-$2107 & NLSy 1 & 0.044 & 8.42	&		2005$-$07$-$26	&	0300240201	&	18.7	\\
KUG 1141+371 & Sy 1.0 & 0.038 & 1.90 & 2009$-$05$-$23 & 0601780501 & 5.4 \\
LEDA 168563		& Sy 1.0 & 0.029 & 54.2 &		2007$-$02$-$26	&	0401790201	&	10.5			\\
MCG $-$02$-$14$-$009 & Sy 1.0 & 0.028 & 9.23 & 2009$-$02$-$27 & 0550640101 & 79.1 \\
MCG$-$02$-$58$-$022	&	Sy 1.5 & 0.047 & 3.60 &	2000$-$12$-$01		&	0109130701	&	10.3			\\
MCG$-$06$-$30$-$015	&	Sy 1.5 & 0.008 & 4.08 &	2001$-$08$-$04	&	0029740801	&	124.0		\\
MCG+08$-$11$-$011	&	Sy 1.5 & 0.020 & 20.9 &	2004$-$04$-$09	&	0201930201	&	30.4			\\
Mrk 1040			&	Sy 1.0 & 0.016 & 7.22 &	2009$-$02$-$13	&	0554990101	&	66.8			\\
Mrk 1044 & NLSy 1 & 0.016 & 3.36 &2013$-$01$-$27 & 0695290101 & 107.2 \\
Mrk 110	& NLSy 1 & 0.035 & 1.47 &		2004$-$11$-$15	&	0201130501	&	47	\\
Mrk 1152			&	Sy 1.5 & 0.052 & 1.68 &	2003$-$06$-$15		&	0147920101	&	21.5			\\
Mrk 279			& Sy 1.0 & 0.030 & 1.78 &	2005$-$11$-$17	&	0302480501	&	49.2			\\
Mrk 290 & Sy 1.5 & 0.030 & 1.70 & 2006$-$05$-$06 & 0400360801 & 18.6 \\
Mrk 335 &NLSy 1 &0.026 & 4.03& 2006$-$01$-$03 & 0306870101 & 117.0 \\
Mrk 352$^P$ & Sy 1.0 & 0.015 & 5.59& 2006$-$01$-$24 & 0312190101 & 10.8 \\
Mrk 359 &NLSy 1 & 0.017 & 4.79 &2010$-$07$-$25 & 0655590201 & 24.8 \\
Mrk 50						&	Sy 1.2 & 0.024 & 1.77 &2010$-$12$-$09 	&	0650590401	&	15.7			\\
Mrk 509						& Sy 1.5 & 0.034 & 4.11 &	2006$-$04$-$25 	&	0306090401	&	61.8			\\
Mrk 590						& Sy 1.0 & 0.027 & 2.68 &	2004$-$07$-$04 	&	0201020201	&	32.4			\\
Mrk 6						&Sy 1.5 & 0.019 & 6.39 &	2001$-$03$-$27	&	0061540101	&	26.0			\\
Mrk 684$^P$ & NLSy 1 & 0.046 & 1.46 & 2006$-$01$-$24 & 0300910101 & 11.1 \\
Mrk 704 & Sy 1.2 & 0.029 & 3.43 & 2008$-$11$-$02 & 0502091601 & 87.0 \\
Mrk 739E & NLSy 1 & 0.029 & 1.90 & 2009$-$06$-$14 & 0601780401 & 10.0 \\
Mrk 766	&NLSy 1 & 0.013 & 1.71 &		2001$-$05$-$20	&	0109141301	&	121.4	\\
Mrk 771 & Sy 1.0 & 0.063& 2.24 &2005$-$07$-$09 & 0301450201 & 24.3 \\
Mrk 79 & Sy 1.2 & 0.022 & 5.73 & 2008$-$04$-$26 & 0502091001 & 67.6 \\
Mrk 817$^P$ & Sy 1.5 & 0.031 & 1.49 & 2009$-$12$-$13 & 0601781401 & 5.6 \\
NGC 3227				&Sy 1.5 & 0.004 & 2.15 &	2006$-$12$-$03 	&	0400270101	&	101.3		\\
NGC 3516					&	Sy 1.5 & 0.009 & 3.05 &2001$-$11$-$09 	&	0107460701	&	119.5		\\
NGC 3783					&	Sy 1.5 & 0.010 & 8.50 &2001$-$12$-$19	&	0112210501	&	127.5		\\
NGC 4051		&	NLSy 1 & 0.002 & 1.32 &2002$-$11$-$22	&	0157560101	&	48.2	\\
NGC 4151					&Sy 1.5 & 0.003 & 1.98 &	2000$-$12$-$22	&	0112830201	&	57.0			\\
NGC 4593					&	Sy 1.0 & 0.009 & 2.31 & 2000$-$07$-$02		&	0109970101	&	10.1			\\
NGC 5548					&	Sy 1.5 & 0.017 & 1.69 &2001$-$07$-$09	&	0089960301	&	70.2			\\
NGC 6814		&	Sy 1.5 & 0.005 & 12.8 &2009$-$04$-$22	&	0550451801	&	28.4	\\
NGC 7469		&	Sy 1.5 & 0.016 & 4.87 &2004$-$11$-$30	&	0207090101	&	84.5	\\
NGC 7603 &Sy 1.5 & 0.030 & 4.09 & 2006$-$06$-$14 & 0305600601 & 16.4 \\
NGC 985	&Sy 1.5 & 0.043 & 2.9&	2003$-$07$-$15	&	0150470601	&	57.6	\\
PG 0804+761	&	Sy 1.0 & 0.100 & 2.98 &	2010$-$03$-$10	&	0605110101	&	42.8	\\
PG 1501+106	&	Sy 1.5 & 0.036 & 2.34 &	2005$-$01$-$16	&	0205340201	&	46	\\
PKS 0558$-$504 & NLSy 1 &0.137 & 4.50 &  2008$-$09$-$11 & 0555170401 & 123.3 \\
PKS 2135$-$14  & Sy 1.5 & 0.200 & 4.70 & 2001$-$04$-$28 & 0092850201 & 27.2 \\
QSO B1419+480	& Sy 1.5 & 0.072 & 1.65 &		2002$-$05$-$27	&	0094740201	&	20.1	\\
QSO B1821+643	&	Sy 1.2 & 0.297 & 4.04 &	2007$-$12$-$10	&	0506210101	&	13.6	\\
RHS 39	&	Sy 1.0 & 0.022 & 4.88 &	2007$-$08$-$05	&	0502090201	&	108.9	\\
RX J2135.9+4728	&	Sy 1.0 & 0.025 & 38.6&	2010$-$11$-$11	&	0650591701	&	22.8	\\
SWIFT J0519.5$-$3140	&	Sy 1.0 & 0.013 & 1.76&	2010$-$01$-$29	&	0610180101	&	76.5	\\
SWIFT J0640.4$-$2554	&	Sy 1.0 & 0.026 & 11.4 &	2006$-$03$-$07	&	0312190801	&	10.3	\\
SWIFT J0917.2$-$6221	&	Sy 1.0 & 0.057 & 19.1 &	2008$-$12$-$14	&	0550452601	&	16	\\
SWIFT J1038.8$-$4942	&	Sy 1.0 & 0.060 & 27.2 &	2011$-$01$-$11	&	0650591101	&	24.9	\\
SWIFT J2009.0$-$6103	&	Sy 1.5 & 0.015 & 4.2 &	2009$-$03$-$30	&	0552170301	&	92.9	\\
UGC 3142	&	Sy 1.0 & 0.022 & 19.0 &	2007$-$03$-$18	&	0401790101	&	9.8	\\

\noalign{\smallskip\hrule\smallskip}
\multicolumn{5}{l}{{\footnotesize $^P$ source presented pile-up and the extraction region used is an annulus of inner radius 15 arcsec}}   \\
\end{longtable}
\end{center}
}


\longtab{3}{
\begin{landscape}
\begin{center}

\begin{longtable}{lccccccccc}

\caption{\textit{EPIC}-\textit{PN} and \textit{MOS} \textit{XMM-Newton} and \textit{Swift}/\textit{BAT} data analysis.}\label{tab:fit}\\
\hline \hline \\

\noalign{\smallskip}
   \multicolumn{1}{l}{Source} &
   \multicolumn{1}{c}{$\Gamma$} &
   \multicolumn{1}{c}{ $F_{\text{continuum}}$ (2-10)} &
   \multicolumn{1}{c}{$F_{\text{soft excess}}$ (0.5-2)} &
   \multicolumn{1}{c}{Model } &
   \multicolumn{1}{c}{$q$} &
\multicolumn{1}{c}{R}  &
\multicolumn{1}{c}{$\chi^{2}$/dof}  \\

\noalign{\smallskip}

  \multicolumn{1}{l}{ } &
   \multicolumn{1}{c}{  } &
   \multicolumn{1}{c}{$10^{-11}\rm\,erg\,cm^{-2}\,s^{-1}$ } &
   \multicolumn{1}{c}{ $10^{-11}\rm\,erg\,cm^{-2}\,s^{-1}$ } &
   \multicolumn{1}{c}{} &
\multicolumn{1}{c}{} &
\multicolumn{1}{c}{} &
   \multicolumn{1}{c}{} \\
\noalign{\smallskip}

\hline
\noalign{\smallskip}

\endfirsthead

\multicolumn{6}{l}%
{\small \hspace{-0.1in}{{\textbf{\tablename\ \thetable{}.} \textit{EPIC}-\textit{PN} and \textit{MOS} \textit{XMM-Newton} and \textit{Swift}/\textit{BAT} data analysis. -- \textit{continued} }}} \\
\noalign{\smallskip}
\hline \hline \\[0.05ex]

\noalign{\smallskip}
   \multicolumn{1}{l}{Source} &
   \multicolumn{1}{c}{$\Gamma$} &
   \multicolumn{1}{c}{  $F_{\text{continuum}}$ (2-10)} &
   \multicolumn{1}{c}{$F_{\text{soft excess}}$ (0.5-2)} &
   \multicolumn{1}{c}{Model }  &
   \multicolumn{1}{c}{$q$} &
\multicolumn{1}{c}{R}  &
\multicolumn{1}{c}{$\chi^{2}$/dof}  \\

\noalign{\smallskip}

  \multicolumn{1}{l}{ } &
   \multicolumn{1}{c}{  } &
   \multicolumn{1}{c}{$10^{-11}\rm\,erg\,cm^{-2}\,s^{-1}$ } &
   \multicolumn{1}{c}{ $10^{-11}\rm\,erg\,cm^{-2}\,s^{-1}$ } &
   \multicolumn{1}{c}{} &
\multicolumn{1}{c}{} &
\multicolumn{1}{c}{} &
  \multicolumn{1}{c}{} \\
\noalign{\smallskip}

\hline 
\noalign{\smallskip}

\endhead

\noalign{\smallskip}
\hline
\endfoot

\endlastfoot

1H 0419$-$577	&$ 1.79\,_{-0.04}^{+0.03}$ & $1.08\,_{-0.002}^{+0.002}$ & $0.43\,_{-0.003}^{+0.003}$ & 2brem & $0.64\,_{-0.005}^{+0.005}$ & $1.11\,_{-0.24}^{+0.20}$ & 4347/3996\\
\noalign{\smallskip}
1H2251$-$179 & $1.57\,_{-0.05}^{+0.05}$ & $1.87\,_{-0.004}^{+0.004}$ & $0.11\,_{-0.003}^{+0.003}$& 1brem + WA & $0.13\,_{-0.004}^{+0.004}$ & $1.15\,_{-0.21}^{+0.24}$ & 3378/3199 \\
\noalign{\smallskip}
1RXSJ213944.3+595016 & $1.69\,_{-0.05}^{+0.09}$ & $0.55\,_{-0.005}^{+0.005}$ & --  & Absorbed & -- & $0.27\,_{-0.14}^{+0.46}$& 654/643\\
\noalign{\smallskip}
2MASSi J1031543$-$141651	&$1.73 \,_{-0.02}^{+0.03}$ & $1.58\,_{-0.004}^{+0.004}$ & $0.79\,_{-0.01}^{+0.01}$ & 2brem & $0.82\,_{-0.011}^{+0.011}$ & $0.38\,_{-0.08}^{+0.09}$ & 3039/3044\\
\noalign{\smallskip}
2MASX J18560128+1538059	&$1.73\,_{-0.07}^{+0.12} $ & $1.03\,_{-0.005}^{+0.005}$ & $0.10\,_{-0.007}^{+0.007}$& Cold abs + 1brem & $0.17\,_{-0.012}^{+0.012}$ & $1.01\,_{-0.33}^{+0.64}$ & 1489/1540\\
\noalign{\smallskip}
2MASX J22484165$-$5109338	&$1.64 \,_{-0.06}^{+0.09}$ & $0.38\,_{+0.0009}^{+0.0009}$ & $0.10\,_{-0.002}^{+0.002}$ & 2brem & $0.53\,_{-0.011}^{+0.011}$ & $0.79\,_{-0.33}^{+0.47}$ & 2202/2195\\
\noalign{\smallskip}
3C 111.0	&$	1.54 \,_{-0.02}^{+0.02}$ & $ 3.92\,_{-0.005}^{+0.005}$ & $0.29\,_{-0.006}^{+0.006}$ & Cold abs + 1brem & $0.17\,_{-0.004}^{+0.004}$ & $0.15\,_{-0.10}^{+0.09}$ & 3430/3147\\
\noalign{\smallskip}
3C 382	&$	1.77 \,_{-0.05}^{+0.06} $& $3.48\,_{-0.006}^{+0.006}$ & $1.14\,_{-0.01}^{+0.01}$ & 2brem + WA & $0.54\,_{-0.005}^{+0.005}$ & $0.32\,_{-0.19}^{+0.18}$ & 2794/2652\\
\noalign{\smallskip}
3C 390.3 &$ 1.57 \,_{-0.02}^{+0.01} $& $3.43\,_{-0.007}^{+0.007}$ &  $0.52\,_{-0.006}^{+0.006}$ & 2brem + WA &$0.32\,_{-0.004}^{+0.004}$& $0.28\,_{-0.007}^{+0.01}$ & 4863/4349\\
\noalign{\smallskip}
4C +74.26&$	1.78\,_{-0.04}^{+0.04}$ & $1.83\,_{-0.003}^{+0.003}$ & -- & Absorbed & --  & $0.99\,_{-0.23}^{+0.23}$ & 4069/3680 \\
\noalign{\smallskip}
4U 0517+17 & $1.67\,_{-0.05}^{+0.05}$ & $2.31\,_{-0.004}^{+0.004}$ & $0.10\,_{-0.002}^{+0.002}$ & 1brem + WA & $0.08\,_{-0.002}^{+0.002}$ & $1.01\,_{-0.19}^{+0.21}$ & 4789/4062\\
\noalign{\smallskip}
6dF J2132022$-$334254 & $1.42 \,_{-0.05}^{+0.13}$ & $0.79\,_{-0.004}^{+0.004}$ & $0.09\,_{-0.003}^{+0.003}$ & 2brem & $0.32\,_{-0.011}^{+0.011}$ & $0.64\,_{-0.21}^{+0.52}$ & 2167/2193\\
\noalign{\smallskip}
Ark 120 & $1.91\,_{-0.02}^{+0.02}$ & $2.97\,_{-0.003}^{+0.003}$ & $1.83\,_{-0.008}^{+0.008}$ &  2brem & $0.82\,_{-0.004}^{+0.004}$ & $0.83\,_{-0.14}^{+0.12}$ & 5219/4654\\
\noalign{\smallskip}
CGCG 229$-$015	& $1.72\,_{-0.12}^{+0.15} $ & $0.32\,_{-0.001}^{+0.001}$ & $0.08\,_{-0.003}^{+0.003}$ & 2brem & $0.44\,_{-0.017}^{+0.017}$ & $0.96\,_{-0.59}^{+0.79}$ & 1520/1567\\
\noalign{\smallskip}
ESO 140$-$43 & $1.97\,_{-0.05}^{+0.04}$ & $1.80\,_{-0.004}^{+0.004}$ & -- & 2brem + WA & -- & $2.15\,_{-0.40}^{+0.28}$ & 2906/2570\\
\noalign{\smallskip}
ESO 141$-$55 & $1.95\,_{-0.02}^{+0.05}$ & $2.82\,_{-0.006}^{+0.006}$ & $0.85\,_{-0.01}^{+0.01}$ & 2brem &  $0.38\,_{-0.005}^{+0.005}$   & $0.71\,_{-0.12}^{+0.44}$ & 3230/2789\\
\noalign{\smallskip}
ESO 198$-$024	& $1.64\,_{-0.02}^{+0.04}$ & $0.81\,_{-0.001}^{+0.001}$ &$ 0.15\,_{-0.002}^{+0.002}$ & 2brem & $0.36\,_{-0.005}^{+0.005}$& $0.38\,_{-0.11}^{+0.22}$ & 2875/2780\\
\noalign{\smallskip}
ESO 209$-$12 & $1.68\,_{-0.05}^{+0.08}$ & $0.68\,_{-0.005}^{+0.005}$ & $0.25\,_{-0.006}^{+0.006}$ & 1brem + WA & $0.68\,_{-0.017}^{+0.017}$ & $1.14\,_{-0.32}^{+0.70}$ & 1075/1034\\
\noalign{\smallskip}
ESO 323$-$77 & -- & -- & -- & Clumpy torus + BLR$^{\text{(a)}}$& -- & -- &--\\
\noalign{\smallskip}
ESO 359$-$ G 019 &$	1.51 \,_{-0.06}^{+0.04}$ & $0.25\,_{-0.002}^{+0.002}$ & $0.05\,_{-0.004}^{+0.004}$ & 2brem & $0.49\,_{-0.039}^{+0.039}$ & $0.79\,_{-0.16}^{+0.53}$ & 1013/1012 \\
\noalign{\smallskip}
ESO 548$-$G081	 & $	1.55\,_{-0.03}^{+0.06} $ & $1.18\,_{-0.008}^{+0.008}$ & $0.50\,_{-0.009}^{+0.009}$ & 1brem + WA & $0.64\,_{-0.012}^{+0.012}$ & $0.33\,_{-0.13}^{+0.16}$ & 1055/1024\\
\noalign{\smallskip}
EXO 055620$-$3820.2 & -- & -- & -- & Ionized absorber$^{\text{(b)}}$ & -- & -- & --\\
\noalign{\smallskip}
Fairall 1116	&$	1.79\,_{-0.04}^{+0.09} $ & $0.60\,_{-0.003}^{+0.003}$ & $0.18\,_{-0.006}^{+0.006}$ & 2brem  & $0.48\,_{-0.016}^{+0.016}$ & $0.41\,_{-0.11}^{+0.54}$ & 1593/1559\\
\noalign{\smallskip}
Fairall 1146 & $2.07\,_{-0.08}^{+0.07}$ & $0.80\,_{-0.006}^{+0.006}$ & -- & 1brem + WA & --  & $3.08\,_{-0.71}^{+0.62}$ & 1633/1583\\
\noalign{\smallskip}
Fairall 9	&$	1.71 \,_{-0.03}^{+0.03}$ & $1.64\,_{-0.002}^{+0.002}$ & $0.32\,_{-0.004}^{+0.004}$ & 2brem & $0.35\,_{-0.004}^{+0.004}$& $0.66\,_{-0.11}^{+0.15}$ & 3396/2997\\
\noalign{\smallskip}
GQ Com	&$	1.62 \,_{-0.08}^{+0.13}$ & $0.41\,_{-0.004}^{+0.004}$ & $0.09\,_{-0.003}^{+0.003}$ & 2brem & $0.45\,_{-0.016}^{+0.016}$ & $0.16\,_{-0.16}^{+0.49}$ & 649/604\\
\noalign{\smallskip}
GRS 1734$-$292 & $1.70\,_{-0.02}^{+0.04}$ & $5.28\,_{-0.02}^{+0.02}$ &  -- & Absorbed  & -- & $0.19\,_{-0.10}^{+0.14}$ & 2033/1971\\
\noalign{\smallskip}
[HB89] 0052+251&$	1.62 \,_{-0.04}^{+0.09}	$ & $0.57\,_{-0.003}^{+0.003}$ & $0.29\,_{-0.007}^{+0.007}$ & 2brem & $1.03\,_{-0.025}^{+0.025}$ & $0.26\,_{-0.21}^{+0.35}$ & 1580/1635\\
\noalign{\smallskip}
[HB89] 0119$-$286	&$1.84 \,_{-0.10}^{+0.07}$ & $0.41\,_{-0.007}^{+0.007}$ &  $0.22\,_{-0.01}^{+0.01}$ & 2brem & $0.79\,_{-0.038}^{+0.038}$ & $2\times10^{-6}\,_{-2\times10^{-6}}^{+0.22}$ & 404/410\\
\noalign{\smallskip}
[HB89] 0241+622 & $ 1.60 \,_{-0.06}^{+0.06}$ & $1.89\,_{-0.004}^{+0.004}$ & $0.17\,_{-0.004}^{+0.004}$ & Cold abs + 1brem & $0.19\,_{-0.005}^{+0.005}$ & $0.95\,_{-0.27}^{+0.27}$ & 3635/3507\\
\noalign{\smallskip}
IC 0486 	& $	1.57 \,_{-0.11}^{+0.16}	$& $0.43\,_{-0.004}^{+0.004}$ & -- & 2WA & -- & $1.47\,_{-0.22}^{1.16}$ & 840/709 \\
\noalign{\smallskip}
IC 2637 	&$	1.62 \,_{-0.08}^{+0.16}$ & $0.36\,_{-0.003}^{+0.003}$ & $0.04\,_{-0.0009}^{+0.0009}$ & 1brem & $0.23\,_{-0.005}^{+0.005}$& $0.52\,_{-0.23}^{1.00}$ & 861/889\\
\noalign{\smallskip}
IC 4329A & $1.69\,_{-0.02}^{+0.02}$ & $8.17\,_{-0.006}^{+0.006}$ & -- &  3WA + Abs + 1brem$^{\text{(c)}}$ & -- & $0.50\,_{-0.08}^{+0.08}$ & 7521/5093\\
\noalign{\smallskip}
IGR J00335+6126 & $1.88\,_{-0.11}^{+0.15}$ & $0.29\,_{-0.003}^{+0.003}$ & -- & Cold abs + 1brem & --  & $5.94\,_{-1.64}^{3.39}$ & 898/778\\
\noalign{\smallskip}
IGR J07597$-$3842 & $1.63\,_{-0.05}^{+0.08}$ & $1.38\,_{-0.006}^{+0.006}$ & $0.11\,_{-0.01}^{+0.01}$ & 1brem + WA & $0.16\,_{-0.02}^{+0.02}$ &$0.82\,_{-0.20}^{+0.44}$ & 2102/2130 \\
\noalign{\smallskip}
IGR J11457$-$1827 & $1.66\,_{-0.01}^{+0.01}$ & $2.30\,_{-0.005}^{+0.005}$ & $1.31\,_{-0.01}^{+0.01}$ & 2brem & $1.09\,_{-0.009}^{+0.009}$ & $0.16\,_{-0.03}^{+0.04}$ & 2736/2533 \\
\noalign{\smallskip}
IGR J12172+0710 & $1.49\,_{-0.11}^{+0.09}$ & $0.53\,_{-0.006}^{+0.006}$ & $0.03\,_{-0.001}^{+0.001}$ & Cold abs + 1brem & $0.14\,_{-0.005}^{+0.005}$ & $0.58\,_{-0.25}^{+0.28}$ & 538/550\\ 
\noalign{\smallskip}
IGR J13038+5348 & $1.58\,_{-0.02}^{+0.02}$ & $1.21\,_{-0.006}^{+0.006}$ & $0.21\,_{-0.009}^{+0.009}$ & 2brem & $0.39\,_{-0.02}^{+0.02}$ & $0.12\,_{-0.05}^{+0.07}$ & 1790/1764\\
\noalign{\smallskip}
IGR J13109$-$5552 & $1.37\,_{-0.06}^{+0.04}$ & $0.47\,_{-0.003}^{+0.003}$ & $0.06\,_{-0.003}^{+0.003}$ & 1brem & $0.39\,_{-0.02}^{+0.02}$ & $4\times10^{-9}\,_{-4\times10^{-9}}^{+0.5}$ & 1167/1159\\
\noalign{\smallskip}
IGR J16119$-$6036 & $1.70\,_{-0.12}^{+0.05}$ & $0.60\,_{-0.004}^{+0.004}$ & $0.09\,_{-0.007}^{+0.007}$ & 2brem & $0.27\,_{-0.02}^{+0.02}$ & $1.21\,_{-0.44}^{+0.32}$ & 1232/1196\\
\noalign{\smallskip}
IGR J16185$-$5928 & $1.49\,_{-0.27}^{+0.10}$ & $0.17\,_{-0.003}^{+0.003}$ & $0.04\,_{-0.002}^{+0.002}$ & 1brem & $0.57\,_{-0.03}^{+0.03}$ & $2.34\,_{-1.35}^{1.63}$ & 410/412\\
\noalign{\smallskip}
IGR J16482$-$3036 & $1.53\,_{-0.03}^{+0.02}$ & $1.61\,_{-0.007}^{+0.007}$ & $0.22\,_{-0.005}^{+0.005}$& Cold abs + 1brem & $0.33\,_{-0.008}^{+0.008}$ & $0.12\,_{-0.06}^{+0.12}$& 2037/1780\\
\noalign{\smallskip}
IGR J16558$-$5203 & $1.64\,_{-0.08}^{+0.11}$ & $1.06\,_{-0.007}^{+0.007}$ & $0.51\,_{-0.02}^{+0.02}$ & 2brem & $0.99\,_{-0.04}^{+0.04}$ & $0.43\,_{-0.24}^{+0.24}$ & 1514/1490\\
\noalign{\smallskip}
IGR J17418$-$1212 & $1.88\,_{-0.03}^{+0.06}$ & $1.48\,_{-0.007}^{+0.007}$ & $0.22\,_{-0.005}^{+0.005}$  & Cold abs + 1brem & $0.21\,_{-0.005}^{+0.005}$ & $0.53\,_{-0.12}^{+0.29}$ & 2112/2047\\
\noalign{\smallskip}
IGR J17488$-$3253 & $1.60\,_{-0.03}^{+0.03}$ & $1.29\,_{-0.009}^{+0.009}$ & $0.05\,_{-0.01}^{+0.01}$ & Cold abs +1brem & $0.08\,_{-0.02}^{+0.02}$ & $2\times10^{-3}\,_{-2\times10^{-3}}^{+0.12}$ & 758/785\\
\noalign{\smallskip}
IGR J18027$-$1455 & $1.49\,_{-0.10}^{+0.14}$ & $0.80\,_{-0.006}^{+0.006}$ & $0.04\,_{-0.003}^{+0.003}$ & Cold abs + 1brem & $0.13\,_{-0.009}^{+0.009}$ & $0.61\,_{-0.39}^{+0.51}$ & 1177/1215\\
\noalign{\smallskip}
IGR J18259$-$0706 & $1.75\,_{-0.06}^{+0.06}$ & $0.70\,_{-0.003}^{+0.003}$ & $0.12\,_{-0.006}^{+0.006}$ & Cold abs + 1brem & $0.29\,_{-0.02}^{+0.02}$ & $1.34\,_{-0.32}^{+0.43}$ & 1926/1872\\
\noalign{\smallskip}
IGR J19378$-$0617 & $2.00\,_{-0.03}^{+0.09}$ & $1.98\,_{-0.009}^{+0.009}$ & $2.45\,_{-0.03}^{+0.03}$ & 2brem + WA & $1.43\,_{-0.02}^{+0.02}$ & $0.43\,_{-0.23}^{+0.92}$ & 1175/1139\\
\noalign{\smallskip}
IGR J21277+5656 & $2.04\,_{-0.03}^{+0.04}$ & $1.81\,_{-0.002}^{+0.002}$ & -- & WA & -- & $1.50\,_{-0.16}^{+0.25}$ & 4992/4378\\
\noalign{\smallskip}
IRAS 04392$-$2713	&$2.06 \,_{-0.20}^{+0.16}	$& $0.49\,_{-0.002}^{+0.002}$ & $0.16\,_{-0.003}^{+0.003}$ & 2brem & $0.51\,_{-0.01}^{+0.01}$ & $3.21\,_{-1.27}^{1.65}$ & 1632/1586\\
\noalign{\smallskip}
IRAS 15091$-$2107 & $1.82\,_{-0.14}^{+0.05}$ & $0.61\,_{-0.003}^{+0.003}$ & $0.02\,_{-0.003}^{+0.003}$ & Cold abs + 1brem & $0.05\,_{-0.007}^{+0.007}$ & $1.90\,_{-0.99}^{+0.59}$ & 1778/1668\\
\noalign{\smallskip}
KUG 1141+371 & $1.49 \,_{-0.12}^{+0.15}$ & $0.21\,_{-0.003}^{+0.003}$ & $0.03\,_{-0.001}^{+0.001}$ & 1brem & $0.36\,_{-0.013}^{+0.013}$& $0.57\,_{-0.53}^{+0.92}$ & 289/315\\
\noalign{\smallskip}
LEDA 168563 & $1.66\,_{-0.01}^{+0.01}$ & $3.03\,_{-0.007}^{+0.007}$ & $2.03\,_{-0.07}^{+0.07}$ & 2brem & $1.32\,_{-0.05}^{+0.05}$ & $2\times 10^{-10}\,_{-2\times 10^{-10}}^{+0.5}$ & 3114/2959\\
\noalign{\smallskip}
MCG $-$02$-$14$-$009 & $1.90 \,_{-0.06}^{+0.08} $ & $0.51\,_{-0.001}^{+0.001}$ & $0.11\,_{-0.003}^{+0.003}$ & 2brem & $0.29\,_{-0.008}^{+0.008}$ & $1.00\,_{-0.25}^{+0.28}$ & 2073/1984\\
\noalign{\smallskip}
MCG $-$02$-$58$-$022 & $1.74\,_{-0.09}^{+0.08}$ & $3.84\,_{-0.009}^{+0.009}$ & $0.26\,_{-0.005}^{+0.005}$ & 1brem & $0.12\,_{-0.002}^{+0.002}$ & $0.44\,_{-0.23}^{+0.22}$ & 1780/1756\\
\noalign{\smallskip}
MCG $-$06$-$30$-$015 & $2.32\,_{-0.05}^{+0.02}$ & $4.69\,_{-0.01}^{+0.01}$ & -- & 3WA + Abs$^{\text{(d)}}$ & -- & $1.05\,_{-0.13}^{+0.09}$& 9970/4823\\
\noalign{\smallskip}
MCG +08$-$11$-$011 & $1.67\,_{-0.04}^{+0.04}$ & $3.48\,_{-0.006}^{+0.006}$ & $0.56\,_{-0.01}^{+0.01}$ & 2brem + WA & $0.31\,_{-0.006}^{+0.006}$ & $1.00\,_{-0.25}^{+0.26}$ & 3139/2907\\
\noalign{\smallskip}
Mrk 1040 & $1.80\,_{-0.04}^{+0.04}$ & $2.09\,_{-0.002}^{+0.002}$ & $0.09\,_{-0.002}^{+0.002}$ & Abs + 2WA + 1brem$^{\text{(e)}}$ & $0.07\,_{-0.002}^{+0.002}$ &$0.97\,_{-0.20}^{+0.22}$ & 4896/4336\\
\noalign{\smallskip}
Mrk 1044	&$	1.99\,_{-0.03}^{+0.04}$ & $0.92\,_{-0.001}^{+0.001}$ & $0.90\,_{-0.006}^{+0.006}$ & 2brem + WA & $1.13\,_{-0.008}^{+0.008}$ & $0.62\,_{-0.11}^{+0.15}$ & 3773/3440\\
\noalign{\smallskip}
Mrk 110 & $1.80\,_{-0.09}^{+0.05}$ & $2.43\,_{-0.006}^{+0.006}$ & $0.68\,_{-0.02}^{+0.02}$ & 2brem & $0.50\,_{-0.02}^{+0.02}$ & $0.74\,_{-0.51}^{+0.31}$ & 2041/2010\\
\noalign{\smallskip}
Mrk 1152 & $1.71\,_{-0.14}^{+0.05}$ & $0.32\,_{-0.002}^{+0.002}$ & $0.06\,_{-0.001}^{+0.001}$ & 1brem + WA &  $0.34\,_{-0.006}^{+0.006}$& $1.01\,_{-0.47}^{+0.17}$ & 1178/1311\\
\noalign{\smallskip}
Mrk 279 & $1.74\,_{-0.01}^{+0.007}$ & $1.69\,_{-0.004}^{+0.004}$ & $0.64\,_{-0.006}^{+0.006}$ & 2brem & $0.70\,_{-0.007}^{+0.007}$ & $0.28\,_{-0.03}^{+0.03}$ & 1564/1557\\
\noalign{\smallskip}
Mrk 290 	&$	1.61\,_{-0.07}^{+0.08}$ & 	$0.52\,_{-0.002}^{+0.002}$ & $0.07\,_{-0.001}^{+0.001}$ & 1brem + WA & $0.28\,_{-0.004}^{+0.004}$ & $1.48\,_{-0.32}^{1.09}$ & 1613/1569\\
\noalign{\smallskip}
Mrk 335	&$	2.06\,_{-0.03}^{+0.05}$ & $1.24\,_{-0.001}^{+0.001}$ & $1.23\,_{-0.007}^{+0.007}$ & 2brem + WA & $1.05\,_{-0.006}^{+0.006}$ & $1.30\,_{-0.26}^{+0.17}$ & 4860/4275\\
\noalign{\smallskip}
Mrk 352	&	$1.44\,_{-0.03}^{+0.09} $ & $0.91\,_{-0.006}^{+0.006}$ & $0.31\,_{-0.009}^{+0.009}$ & 2brem & $0.92\,_{-0.027}^{+0.027}$& $0.04\,_{-0.04}^{+0.33}$ & 1498/1459\\
\noalign{\smallskip}
Mrk 359	&$	1.85\,_{-0.10}^{+0.12} $&$ 0.60\,_{-0.003}^{+0.003} $& $0.18\,_{-0.006}^{+0.006}$ & 2brem & $0.44\,_{-0.02}^{+0.02}$ & $1.11\,_{-0.40}^{+0.72}$ & 1313/1307\\
\noalign{\smallskip}
Mrk 50 & $1.92\,_{-0.13}^{+0.07}$ & $0.70\,_{-0.002}^{+0.002}$ & $0.06\,_{-0.004}^{+0.004}$ & 2brem  & $0.12\,_{-0.008}^{+0.008}$ & $1.75\,_{-0.67}^{+0.55}$ & 1993/1857\\
\noalign{\smallskip}
Mrk 509 & $1.72\,_{-0.04}^{+0.04}$ & $3.58\,_{-0.004}^{+0.004}$ & $1.22\,_{-0.01}^{+0.01}$ & 2brem + WA &  $0.60\,_{-0.005}^{+0.005}$  & $0.43\,_{-0.18}^{+0.17}$ & 4846/4325\\
\noalign{\smallskip}
Mrk 590 & $1.57\,_{-0.03}^{+0.03}$ & $0.49\,_{-0.001}^{+0.001}$ & $0.09\,_{-0.001}^{+0.001}$ & 1brem & $0.41\,_{-0.005}^{+0.005}$ & $0.49\,_{-0.06}^{+0.10}$ & 2614/2513\\
\noalign{\smallskip}
Mrk 6 & $1.44\,_{-0.04}^{+0.14}$ & $1.16\,_{-0.008}^{+0.008}$ & -- & 2WA + Abs + 1brem$^{\text{(f)}}$  & -- & $0.72\,_{-0.24}^{+0.46}$ &1095/1058\\
\noalign{\smallskip}
Mrk 684	&$	1.74\,_{-0.09}^{+0.11} $ & $0.36\,_{-0.003}^{+0.003}$ & $0.15\,_{-0.005}^{+0.005}$ & 2brem & $0.72\,_{-0.03}^{+0.03}$& $0.06\,_{-0.06}^{+0.35}$ & 803/848\\
\noalign{\smallskip}
Mrk 704	&$	1.94 \,_{-0.06}^{+0.05}$ & $1.24\,_{-0.002}^{+0.002}$ & $0.13\,_{-0.009}^{+0.009}$ & 1brem + 2WA & $0.13\,_{-0.009}^{+0.009}$ & $1.08\,_{-0.20}^{+0.17}$ & 4115/3649\\
\noalign{\smallskip}
Mrk 739E	&$	1.96 \,_{-0.09}^{+0.14} $& $0.67\,_{-0.003}^{+0.003}$ & $0.02\,_{-0.002}^{+0.002}$ & Cold abs + 1brem & $0.04\,_{-0.004}^{+0.004}$ & $0.72\,_{-0.34}^{+0.59}$ & 1121/1171\\
\noalign{\smallskip}
Mrk 766 & $2.02\,_{-0.02}^{+0.02}$ & $1.70\,_{-0.002}^{+0.002}$ & $1.25\,_{-0.008}^{+0.008}$  & 2WA + 2brem &   $0.83\,_{-0.005}^{+0.005}$ &$0.95\,_{-0.11}^{+0.13}$& 4276/2981\\
\noalign{\smallskip}
Mrk 771	& $	1.79 \,_{-0.04}^{+0.09} $& $0.42\,_{-0.003}^{+0.003}$ & $0.15\,_{-0.003}^{+0.003}$ & 2brem & $0.57\,_{-0.01}^{+0.01}$ & $0.26\,_{-0.07}^{+0.30}$ & 1379/1338\\
\noalign{\smallskip}
Mrk 79 	&$	1.77 \,_{-0.07}^{+0.06}$ & $0.32\,_{-0.001}^{+0.001}$ & $0.06\,_{-0.009}^{+0.009}$ & 1brem + WA & $0.31\,_{-0.05}^{+0.05}$ & $8.29\,_{-2.92}^{1.98}$& 2309/2226\\
\noalign{\smallskip}
Mrk 817 	&$	2.00 \,_{-0.15}^{+0.14}$ & $0.97\,_{-0.009}^{+0.009}$ & $0.51\,_{-0.02}^{+0.02}$ & 2brem & $0.61\,_{-0.03}^{+0.03}$ & $1.86\,_{-0.94}^{1.07}$ & 750/733\\
\noalign{\smallskip}
NGC 3227 & $1.58\,_{-0.02}^{+0.02}$ & $2.97\,_{-0.003}^{+0.003}$ & -- & 3WA + 2brem$^{\text{(g)}}$ & -- & $0.70\,_{-0.08}^{+0.10}$ & 5787/4816\\
\noalign{\smallskip}
NGC 3516 & $1.79\,_{-0.02}^{+0.01}$ & $1.91\,_{-0.009}^{+0.009}$ &-- & 3WA$^{\text{(h)}}$ & --& $3.21\,_{-0.19}^{+0.31}$ & 8131/4424\\
\noalign{\smallskip}
NGC 3783 & $1.64\,_{-0.03}^{+0.02}$ & $4.06\,_{-0.005}^{+0.005}$ &-- & 2WA + 2brem$^{\text{(i)}}$ & --& $1.24\,_{-0.14}^{+0.08}$ & 2332/1896\\
\noalign{\smallskip}
NGC 4051 & $1.77\,_{-0.05}^{+0.08}$ & $0.84\,_{-0.002}^{+0.002}$ & --& WA + 1brem$^{(j)}$ & -- & $1.89\,_{-0.17}^{+0.12}$  & 7602/2576 \\
\noalign{\smallskip}
NGC 4151 & $1.58\,_{-0.002}^{+0.01}$ & $6.70\,_{-0.02}^{+0.02}$ & -- &  2WA + Abs$^{(k)}$ & -- & $1.38\,_{-0.07}^{+0.04}$ & 19675/4849\\
\noalign{\smallskip}
NGC 4593 & $1.71\,_{-0.01}^{+0.03}$ & $2.73\,_{-0.01}^{+0.01}$ & $1.22\,_{-0.02}^{+0.02}$ & 2brem + WA & $0.80\,_{-0.01}^{+0.01}$ & $0.45\,_{-0.06}^{+0.09}$ & 1191/1195\\
\noalign{\smallskip}
NGC 5548 & $1.62\,_{-0.009}^{+0.005}$ & $2.92\,_{-0.003}^{+0.003}$ & $0.64\,_{-0.006}^{+0.006}$ & 2brem + 2WA & $0.45\,_{-0.004}^{+0.004}$ & $0.25\,_{-0.02}^{+0.02}$ & 3526/3230\\
\noalign{\smallskip}
NGC 6814 & $1.71\,_{-0.09}^{+0.09}$ & $1.90\,_{-0.004}^{+0.004}$ & $0.22\,_{-0.01}^{+0.01}$ & 2brem  & $0.20\,_{-0.009}^{+0.009}$ & $0.92\,_{-0.37}^{+0.40}$ & 2641/2453\\
\noalign{\smallskip}
NGC 7469 & $1.88\,_{-0.04}^{+0.04}$ & $2.42\,_{-0.003}^{+0.003}$ &  $1.26\,_{-0.01}^{+0.01}$ & 2brem + WA & $0.72\,_{-0.006}^{+0.006}$& $0.98\,_{-0.23}^{+0.26}$ & 1701/1734 \\
\noalign{\smallskip}
NGC 7603&$	1.82\,_{-0.07}^{+0.10} $ & $2.11\,_{-0.01}^{+0.01}$ & $0.78\,_{-0.02}^{+0.02}$ & 2brem & $0.56\,_{-0.02}^{+0.02}$& $0.39\,_{-0.22}^{+0.37}$ & 1035/1007\\
\noalign{\smallskip}
NGC 985 & $1.65\,_{-0.05}^{+0.08}$ & $0.82\,_{-0.004}^{+0.004}$ & -- & 1brem + WA & --  & $1.36\,_{-0.22}^{+0.51}$ & 3034/2710\\
\noalign{\smallskip}
PG 0804+761 & $1.96\,_{-0.04}^{+0.09}$ & $1.11\,_{-0.005}^{+0.005}$ & $0.57\,_{-0.008}^{+0.008}$ & 2brem & $0.64\,_{-0.009}^{+0.009}$ & $0.48\,_{-0.15}^{+0.49}$ & 1841/1784\\
\noalign{\smallskip}
PG 1501+106 & $1.61\,_{-0.03}^{+0.11}$ & $0.80\,_{-0.004}^{+0.004}$  & $0.23\,_{-0.007}^{+0.007}$ & 2brem + WA & $0.61\,_{-0.02}^{+0.02}$ & $1.73\,_{-0.27}^{1.02}$ & 2022/1998\\
\noalign{\smallskip}
PKS 0558$-$504	& $ 1.99\,_{-0.02}^{+0.04} $ & $1.15\,_{-0.002}^{+0.002}$ & $1.05\,_{-0.006}^{+0.006}$ & 2brem & $1.08\,_{-0.006}^{+0.006}$ & $0.26\,_{-0.14}^{+0.21}$ & 1864/1626\\
\noalign{\smallskip}
PKS 2135$-$14	&$1.57\,_{-0.05}^{+0.04}$ &$0.43\,_{-0.002}^{+0.002}$ & $0.08\,_{-0.002}^{+0.002}$ &2brem & $0.41\,_{-0.01}^{+0.01}$ & $0.29\,_{-0.08}^{+0.20}$ & 1889/1964\\
\noalign{\smallskip}
QSO B1419+480 & $1.73\,_{-0.05}^{+0.24}$ & $0.52\,_{-0.004}^{+0.004}$ & $0.14\,_{-0.003}^{+0.003}$ & 1brem + WA & $0.39\,_{-0.009}^{+0.009}$ & $0.56\,_{-0.17}^{2.24}$ & 1333/1274\\
\noalign{\smallskip}
QSO B1821+643 & $1.94\,_{-0.11}^{+0.22}$ & $0.96\,_{-0.004}^{+0.004}$ & $0.34\,_{-0.01}^{+0.01}$ & 2brem & $0.45\,_{-0.01}^{+0.01}$ & $1.79\,_{-0.76}^{2.87}$ & 1830/1762\\
\noalign{\smallskip}
RHS 39 & $1.80\,_{-0.04}^{+0.03}$ & $1.37\,_{-0.002}^{+0.002}$ & $0.59\,_{-0.004}^{+0.004}$ & 2brem & $0.68\,_{-0.005}^{+0.005}$  & $1.10\,_{-0.22}^{+0.16}$ & 4788/4368\\
\noalign{\smallskip}
RX J2135.9+4728 & $1.88\,_{-0.10}^{+0.16}$ & $0.51\,_{-0.002}^{+0.002}$ & $0.01\,_{-0.004}^{+0.004}$ & 1brem + WA & $0.03\,_{-0.01}^{+0.01}$ &  $3.45\,_{-0.73}^{1.43}$ & 1962/1893\\
\noalign{\smallskip}
SWIFT J0519.5$-$3140 & $1.76\,_{-0.04}^{+0.02}$ & $0.47\,_{-0.001}^{+0.001}$ & $0.02\,_{-0.007}^{+0.007}$ & 1brem + WA & $0.03\,_{-0.01}^{+0.01}$ & $4.83\,_{-0.27}^{+0.26}$ & 3845/3582\\
\noalign{\smallskip}
SWIFT J0640.4$-$2554 & $1.61\,_{-0.03}^{+0.03}$ & $1.94\,_{-0.009}^{+0.009}$ & -- & Absorbed & -- & $0.27\,_{-0.27}^{+0.50}$& 2261/2120\\
\noalign{\smallskip}
SWIFT J0917.2$-$6221 & $1.88\,_{-0.07}^{+0.06}$ & $1.25\,_{-0.01}^{+0.01}$  & -- & 2WA + Abs + 2brem &  -- &$0.60\,_{-0.15}^{+0.23}$& 1194/1039\\
\noalign{\smallskip}
SWIFT J1038.8$-$4942 & $1.58\,_{-0.17}^{+0.05}$ & $0.64\,_{-0.01}^{+0.01}$ & --& 2WA + Abs + 2brem &  --&$0.99\,_{-1.86}^{+0.54}$& 750/576\\
\noalign{\smallskip}
SWIFT J2009.0$-$6103 & $1.77\,_{-0.05}^{+0.04}$ & $1.56\,_{-0.004}^{+0.004}$  &-- & 2WA &  --&$0.98\,_{-0.22}^{+0.18}$& 7232/4266\\
\noalign{\smallskip}
UGC 3142 & $1.41\,_{-0.03}^{+0.02}$ & $2.18\,_{-0.02}^{+0.02}$ & -- & 2WA + Abs + 3brem$^{(l)}$  & -- &$0.61\,_{-0.06}^{+0.05}$& 1543/1223\\
\noalign{\smallskip}

\noalign{\smallskip\hrule\smallskip}
\multicolumn{7}{l}{{\footnotesize Models: \textit{brem}: Bremsstrahlung models; \textit{WA}: warm absorption; \textit{Abs/Cold abs}: cold absorption.}}   \\
\multicolumn{7}{l}{{\footnotesize References: (a) \cite{Miniutti2014}; (b) \cite{Turner1996}; (c) \cite{Steenbrugge2005}; (d) \cite{Miyakawa2012}; (e) \cite{Reynolds1995};}}\\
\multicolumn{7}{l}{{\footnotesize (f) \cite{Mingo2011}; (g) \cite{Noda2014}; (h) \cite{Huerta2014}; (i) \cite{Brenneman2011}; (j) \cite{Pounds2013}; (k) \cite{Wang2011}; (l) \cite{Ricci2010}.}}   \\
\end{longtable}
\end{center}
\end{landscape}
}


\begin{acknowledgements}
We would like to thank the referee for her useful comments which helped to improve this paper.
We also acknowledge D. Eckert for useful discussion and L. Gibaud for her help in reducing data.
RB acknowledges a grant from the Swiss National Science Foundation. 
CR acknowledges financial support from the Japan Society for the Promotion of Science (JSPS), CONICYT-Chile ``EMBIGGEN" Anillo (grant ACT1101), from FONDECYT 1141218 and Basal-CATA PFB--06/2007. 
This research has made use of the NASA/IPAC Extragalactic Database (NED) which is operated by the Jet Propulsion Laboratory, of data obtained from the High Energy Astrophysics Science Archive Research Center (HEASARC), provided by NASA's Goddard Space Flight Center and of the SIMBAD Astronomical Database which is operated by the Centre de Donn\'ees astronomiques de Strasbourg. This work was based on observations obtained with \textit{XMM-Newton}, an ESA science mission with instruments and contributions directly funded by ESA Member States and NASA. 
\end{acknowledgements}

\bibliographystyle{aa}
\bibliography{soft_excess.bib}

\end{document}